\documentclass[12pt,preprint]{aastex}
\def\lta{\lower2pt\hbox{$\buildrel {\scriptstyle <} 
   \over {\scriptstyle\sim}$}}
\def\gta{\lower2pt\hbox{$\buildrel {\scriptstyle >} 
   \over {\scriptstyle\sim}$}}

\def\fop{f_{op}}

\begin{document}

\title{Gamma-ray Burst 080319B: Evidence for Relativistic Turbulence, Not
   Internal Shocks}

\author{Pawan Kumar$^1$ \& Ramesh Narayan$^2$ \\
$^1$Astronomy Department, University of Texas, Austin, TX 78712 \\
$^2$Harvard-Smithsonian Center for Astrophysics, 60 Garden Street, Cambridge, MA 02138\\
}

\begin{abstract}

We show that the excellent optical and gamma-ray data available for
GRB~080319B rule out the internal shock model for the prompt emission.
The data instead point to a model in which the observed radiation was
produced close to the deceleration radius ($\sim10^{17}$\,cm) by a
turbulent source with random Lorentz factors $\sim10$ in the comoving
frame. The optical radiation was produced by synchrotron emission
from relativistic electrons, and the gamma-rays by inverse Compton
scattering of the synchrotron photons.  The gamma-ray emission
originated both in eddies and in an inter-eddy medium, whereas the
optical radiation was mostly from the latter.  Therefore, the
gamma-ray emission was highly variable whereas the optical was much
less variable.  The model explains all the observed features in the
prompt optical and gamma-ray data of GRB~080319B. We are unable to
determine with confidence whether the energy of the explosion was
carried outward primarily by particles (kinetic energy) or magnetic
fields.  Consequently, we cannot tell whether the turbulent medium was
located in the reverse shock (we can rule out the forward shock) or in
a Poynting-dominated jet. 

\end{abstract}


\section{Introduction}

Two major unsolved questions in the field of gamma-ray bursts are: (i)
the mechanism by which the energy in relativistic jets is converted to
random particle kinetic energy, and (ii) the radiation process by
which the particle energy is converted to gamma-ray photons.

A number of ideas have been proposed for the conversion of jet energy
to particle energy (cf. Piran 1999, 2005; M{\'e}sz{\'a}ros 2002;
Thompson 1994, 2006; Lyutikov \& Blandford 2003; Zhang 2007).  Most
popular among these is the so-called internal shock model (Piran,
Shemi \& Narayan 1993; Rees \& Meszaros 1994; Katz 1994), in which
different parts of the relativistic GRB jet travel at different
speeds.  Faster segments collide with slower segments in shocks, and a
fraction of the jet kinetic energy is converted to thermal energy.
Gamma-rays are then produced by either the synchrotron process or the
synchrotron-self-Compton (SSC) process (e.g., Piran 2005,
M{\'e}sz{\'a}ros 2002, Zhang 2007).  Another model is the external
shock model (Dermer 1999) in which gamma-rays are produced via the
synchrotron process in an external shock driven into the circumstellar
medium by the GRB jet.  Difficulties with this model have been pointed
out by a number of authors (cf. Piran 1999).

The excellent data obtained by the Swift and Konus satellites for GRB
080319B -- dubbed the ``the naked eye burst" -- has provided a new
opportunity to investigate the viability of these models and to
understand the fundamental nature of GRBs.  We summarize here the main
observational properties of this burst (details may be found in
Racusin et al.  2008).  

GRB 080319B lasted for about 50~s and had a burst fluence in the
20~keV~--~7~MeV band of $5.7 \pm 0.1 \times 10^{-4}$~erg~cm$^{-2}$,
which corresponds to an isotropic energy release of $E_\gamma = 1.3
\times 10^{54}$~erg (Golenetskii et al. 2008) for a redshift $z=0.937$
(Vreeswijk et al 2008; Cucchiara \& Fox 2008).  The time-averaged
gamma-ray spectrum during the burst had a peak at around 650~keV.  The
maximum flux at this energy was $\sim7$~mJy, and the time averaged
flux was $\sim 3$~mJy.  The time-averaged gamma-ray spectrum was
measured by Konus-Wind (Racusin et al. 2008) to be $F_\nu \propto
\nu^{0.18 \pm 0.01}$ for photon energies below 650~keV and $F_\nu
\propto \nu^{-2.87 \pm 0.44}$ at higher energies.  (The spectrum
evolved during the course of the burst, as we discuss in \S2, and this
provides additional information on the radiation process.)  In the
optical band, at photon energies around 2~eV, the peak flux of GRB
080319B was $V=5.4$~mag or 20~Jy, and the time-averaged flux was about
10~Jy (Karpov et al 2008). The optical lightcurve varied on a time
scale of about 5~s, while the $\gamma$-ray flux varied on time scales
of $\sim0.5$~s.

GRB 080319B has seriously challenged our understanding of gamma-ray
bursts.  The problems posed by this burst are described in Kumar \&
Panaitescu (2008) and discussed in further detail in the present
paper.  We show that the observations cannot be explained with any of
the standard versions of the internal shock model.  We find, however,
that a consistent model is possible if we give up the idea of internal
shocks and instead postulate that the gamma-ray source is
relativistically turbulent (see Narayan \& Kumar 2008).

We begin in \S2 by arguing that the radiation in GRB 080319B must have
been produced by the SSC mechanism.  Following this, we derive in \S3
the basic equations describing a GRB that radiates via SSC.  In \S4,
we combine these equations with the internal shock model and attempt
to explain the data on GRB 080319B.  We find that no consistent model
is possible.  In \S5, we consider a relativistically turbulent model,
again with SSC radiation, and show that in this case it is possible to
obtain a consistent model of GRB 080319B.  We summarize the main
conclusions in \S6. The Appendix discusses the effects of source
inhomogeneity on our calculation of the synchrotron self-absorption
frequency, and the synchrotron and inverse-Compton fluxes.

\section{Why synchrotron-self-Compton model?}

Racusin et al. (2008; Fig. 3) and Wozniak et al. (2008; Fig. 4) show
that the $\gamma$-ray and optical lightcurves (LCs) of GRB 080319B
have a similar general shape (although the $\gamma$-ray LC is a lot
more variable than the optical LC).  This suggests that the optical
and $\gamma$-ray radiation were produced by the same source.  An
independent theoretical argument in support of this conclusion is
given in \S4.1.  The radiation mechanisms in the two bands must,
however, be different since the optical flux is larger by a factor
$\sim10^4$ than the $\gamma$-ray flux extrapolated to the optical
band.


The average spectral properties of GRB 080319B were summarized in \S1.
Racusin et al. (2008; supplementary material -- Table 1) also reported
spectral fits corresponding to three independent time segments of the
burst: $-2$~s to 8~s, 12~s to 22~s, 26~s to 36~s (all times measured
with respect to the nominal start time of the burst).  The peak of the
gamma-ray spectrum evolved from about 750~keV early in the burst to
about 550~keV at late times, giving a mean peak energy of 650~keV as
mentioned earlier.  More interestingly, the spectral slope at energies
below the peak evolved with time: $F_\nu \propto \nu^{0.50\pm0.04},
\nu^{0.17\pm0.02}, \nu^{0.10\pm0.03}$, during the three time segments.
The unusually hard spectrum during the first time segment
unambiguously points to an SSC origin for the gamma-ray emission, as
we now argue.

The hardest spectrum possible with optically thin synchrotron emission
is $F_\nu\propto \nu^{1/3}$.  The only way to obtain a harder spectrum
is to invoke self-absorption, in which case the spectrum will switch
to $F_\nu \propto \nu^2$ below the self-absorption break.  However, in
order to obtain a synchrotron spectrum with a mean spectral index of
$0.5$ in the band between 20~keV and 650~keV, we would need to have
the self-absorption break at an energy $\sim50-100$~keV.  This has two
serious problems.  First, it is virtually impossible to push the
self-absorption break to such a large energy with any reasonable
parameters for the radiating medium\footnote{For the synchrotron
self-absorption frequency to be $\sim 50$~keV and the flux at 650~keV
to be $\sim 10$~mJy, the distance of the source from the center of the
explosion must be less than 10$^8$~cm. At such a small radius the
medium would be extremely opaque to Thomson scattering and
$\gamma+\gamma\rightarrow e^+ + e^-$.  Therefore, the emergent
radiation would be thermal and no photons with energy $\gta$~1~MeV
would be able to escape from the source.  However, Konus-Wind detected
$\sim 10$~MeV photons from GRB 080319B.}.  Second, a spectral break in
which the slope changes from 1/3 to 2 would almost certainly be
detectable in the data and would not be consistent with a single
power-law with a slope of $0.5$.

A Comptonization model gets around these difficulties.  If the
gamma-ray emission is produced by Compton-scattering, then any break
in the spectrum is not intrinsic to the gamma-rays but merely a
reflection of a break in the spectrum of the underlying soft photons.
If the soft photons are in the optical-infrared band and are produced
by synchrotron emission, then the self-absorption break must be around
0.1~eV, which is perfectly compatible with reasonable model
parameters.  In addition, although the synchrotron spectrum below the
break would be very hard, viz., $F_\nu \propto \nu^2$, the
corresponding segment in the up-scattered inverse-Compton radiation
would be softer: $F_\nu \propto \nu$ (Rybicki \& Lightman, 1979). Thus, 
the gamma-ray spectrum would break from a slope of 1/3 to 1, which is
consistent with the observations, especially when we allow for a
smooth rollover from one spectral slope to the other over a range of
energies.

Why did the gamma-ray spectral slope below 650~keV switch to
$\sim0.1-0.2$ at later times?  The likely explanation is that the
self-absorption frequency of the synchrotron emission dropped to yet
lower energies (in the infrared), and so the break in the gamma-ray
band was pushed closer to, or even below, 20~keV.

In this discussion, we have assumed that the soft radiation is
produced by the synchrotron process.  The alternative is thermal
radiation, but this can be ruled out as it requires a Lorentz factor
of $\sim 10^8 (\delta t)^{-1} T_5^{-1/2}$ to explain the observed flux
of 10 Jy
\footnote{For a relativistically moving thermal source with a
temperature $T$ (in the observer frame) and radius $R$, the observed
flux in the optical band at a frequency
$\nu_{op}\sim5\times10^{14}$~Hz is: $f_{op} \approx 2 k T
(1+z)^4\nu_{op}^2 R^2/(d_L^2 c^2 \Gamma^2)\approx 2 k T (1+z)^2
\nu_{op}^2 (\delta t)^2 \Gamma^2/d_L^2$. Therefore, to explain the
observed optical flux of 10 Jy we require $\Gamma\sim 10^8 (\delta
t)^{-1}T_5^{-1/2}$.}; here $\delta t$ is the observed variability time
(in seconds) of the optical lightcurve and $T_5=T/10^5$K is the
temperature of the source.  The parameter $T_5$ cannot be much larger
than unity since the total energy release would become excessive
($>10^{55}$~erg).  A Lorentz factor of $10^8$ is not reasonable
either.  Therefore, a thermal model for the optical emission is ruled
out.

We thus conclude that the optical photons in GRB 0803019B were
produced by the synchrotron mechanism, and the gamma-ray photons were
produced by the same relativistic electrons by inverse-Compton
scattering the synchrotron photons.  That is, all the observed
radiation in GRB 080319B was the result of the SSC process (\S4 gives
a more detailed discussion).

\section{Synchrotron and SSC processes for a relativistic transient source:
   basic equations}

We first determine what properties the source of optical emission in
GRB 080319B must have, assuming that the radiation is produced by
synchrotron emission (\S3.1). We then consider the gamma-ray data and
describe the additional constraints they provide (\S3.2).  Detailed
application to GRB 080319B is discussed in \S4 and \S5.

\subsection{Modeling the prompt optical data}

The properties of a synchrotron source can be described by five
parameters: $B$, $N_e$, $\Gamma$, $\gamma_i$ and $\tau_e$, which are
the magnetic field strength, the total number of electrons, the bulk
Lorentz factor of the source with respect to the GRB host galaxy, the
{\it typical} electron Lorentz factor in the comoving frame of the
source (the electron distribution is $dn_e/d\gamma\propto \gamma^{-p}$
for $\gamma>\gamma_i$), and the optical depth of the source to Thomson
scattering\footnote{Technically, the electron index $p$ is a sixth
parameter, but since it can be estimated directly from the high energy
slope of the gamma-ray spectrum we do not count it}.  If the source is
at redshift $z$, the peak frequency $\nu_i$ of the synchrotron
spectrum and the observed flux $f_i$ at the peak, both as seen by the
observer, are given by (Rybicki \& Lightman, 1979)
\begin{eqnarray}
&& \nu_i={\phi_\nu(p) qB\gamma_i^2\Gamma \over 2\pi m_e c(1+z)}=
   (1.15\times10^{-8}\,{\rm eV})\, \phi_\nu B\gamma_i^2\Gamma (1+z)^{-1} , \\
&& f_i = { \sqrt{3} \phi_f(p) q^3 B N_e\Gamma (1+z)\over 4\pi d_L^2 m_e c^2}=
   (0.18\,{\rm Jy})\, \phi_f N_{e55} B\Gamma (1+z) d_{L28}^{-2}, 
\label{nui}
\end{eqnarray}
where $d_L$ is the luminosity distance\footnote{The factor (1+z) in
the numerator of the expression for flux, eq. 2, is due to the fact
that the luminosity distance refers to the bolometric flux whereas we
are considering the flux per unit frequency.} to the source, and
$\phi_\nu$ and $\phi_f$ are dimensionless constants that depend on the
electron energy distribution index $p$; for $p=5$ (as suggested by the
high energy spectral index for GRB 080319B), $\phi_\nu=0.5$ and
$\phi_f=0.7$ (cf. Wijers \& Galama, 1999\footnote{$\phi_\nu = 1.5 x_p$
and $\phi_f=\phi_p$ in the notation of Wijers \& Galama.}). In all
subsequent equations we explicitly use these values for the $\phi$'s
with the exception that for the most important quantities -- such as
the total energy and the inverse Compton flux -- we show the
dependence of the results on the $\phi$'s to indicate how
uncertainties in $\nu_i$ and $f_i$ affect the final result.

Throughout the paper we measure frequencies in eV and fluxes in Jy.
All other quantities are in cgs units, but we use the short hand
notation $x_n\equiv x/(10^{n}\,{\rm cgs})$ to scale numerical values,
e.g., $R_{15}=R/(10^{15}\,{\rm cm})$ is the radius of the radiating
shell with respect to the center of the explosion in units of
$10^{15}\,{\rm cm}$.  Using this convention, the source optical depth
$\tau_e$ and the duration of a pulse in the lightcurve $\delta t$ (in
seconds) are given by:
\begin{eqnarray}
&&  \tau_e = {\sigma_T N_e \over 4\pi R^2} = 0.5 {N_{e55}\over R_{15}^2} \quad
   \quad {\rm or} \quad\quad N_{e55} = 2\tau_e R_{15}^2, \label{ne1}\\
&& \delta t = {(1+z) R\over 2 c \Gamma^2} = (1.7\times10^4\,{\rm s}) R_{15}
    \Gamma^{-2}(1+z). 
\label{tau}
\end{eqnarray}
Equivalently,
\begin{equation}
 \Gamma = 130  (\delta t)^{-1/2} (1+z)^{1/2} R_{15}^{1/2}.
\label{gam1}
\end{equation}

We can combine equations (1) \& (2) to eliminate $B$:
\begin{equation}
f_i = {3\times10^7{\rm Jy}\over \gamma_i^2} \nu_i N_{e55} (1+z)^2  
   d_{L28}^{-2},
\end{equation}
and use equation (3) to replace $N_e$ by $\tau_e$. The resultant
equation is:
\begin{equation}
f_i = {4.2\times10^7 {\rm Jy} \over \gamma_i^4} \nu_i Y R_{15}^2 (1+z)^2 
    d_{L28}^{-2},
\end{equation}
or
\begin{equation}
\gamma_i = 81 f_i^{-1/4} \nu_i^{1/4} Y^{1/4} R_{15}^{1/2} (1+z)^{1/2}
  d_{L28}^{-1/2}.
\label{gami1}
\end{equation}
Here we have defined the quantity
\begin{equation}
Y\equiv \gamma_i^2\tau_e,
\label{Ypar}
\end{equation}
which is closely related to, but smaller by a factor of 2-3 compared with,
 the Compton-Y parameter for
GRB 080319B (the difference depends on electron distribution function).
  Substituting eqs. (5) \&
(\ref{gami1}) back into eq. (1), we find
\begin{equation}
B = (205 G) (\delta t)^{1/2} \nu_i^{1/2} f_i^{1/2} Y^{-1/2} R_{15}^{-3/2}
    (1+z)^{-1/2} d_{L28},
\label{b1}
\end{equation}
while equations (\ref{ne1}) \& (\ref{gami1}) yield
\begin{equation}
N_{e55} = 2.9\times10^{-4} \nu_i^{-1/2} f_i^{1/2} Y^{1/2} R_{15}
    (1+z)^{-1} d_{L28}.
\label{ne2}
\end{equation}

Equations (\ref{ne1}), (\ref{gam1}), (\ref{gami1}), (\ref{b1}) and
(\ref{ne2}) are solutions for the five basic physical parameters of
the synchrotron source in terms of the observed variability time
$\delta t$ and four unknown quantities: the source radius $R_{15}$,
the Compton parameter $Y$, the peak energy of the synchrotron spectrum
$\nu_i$ and the synchrotron flux at the peak $f_i$.  The last two
quantities are not independent --- they are constrained by the
observed optical flux $\fop$ ($=10$~Jy in the case of GRB 080319B) at
2 eV.  Depending on whether the synchrotron peak frequency $\nu_i$ is
below or above 2~eV, we obtain the following constraint:
\begin{equation}
\fop = \left\{ \begin{array}{ll}
\hskip -7pt f_i (2/\nu_i)^{1/3}, \quad\quad\quad\quad & \nu_i>2, \\
   & \\
\hskip -7pt f_i (\nu_i/2)^{p/2},
   & \nu_i<2,
\end{array} \right.
\label{fop0}
\end{equation}
where we have used standard results for the spectral slope below and
above the synchrotron peak, assuming for the latter that the cooling
frequency is close to $\nu_i$, as suggested by the spectrum of GRB
080319B\footnote{The spectrum of GRB 080319B peaked at 650 keV and
contained no other break between 20keV and 7 MeV. Since the cooling
frequency cannot be smaller than 650 keV since the spectrum varies as
$\nu^{0.2}$ below the peak, it must be either larger than 7MeV or
close to 650 keV. The former possibility is unlikely since it
corresponds to a radiatively inefficient system and would increase the
energy requirement of an already extreme burst.}; for other bursts,
where the cooling break is substantially above the synchrotron peak,
the equations in this paper can be easily modified by replacing $p$
with $(p-1)$.

Another constraint is provided by the synchrotron self-absorption
frequency $\nu_a$, which can be estimated by equating the synchrotron
flux at $\nu_a$ to the blackbody flux in the Rayleigh-Jeans limit.
Assuming that the electrons that dominate at $\nu_a$ have Lorentz
factor $\gamma_i$, we obtain
\begin{equation}
{2\nu_a{'^2}\over c^2} m_e c^2 \gamma_i = f'(\nu_a') = {f_i\over \Gamma}
   {d_L^2\over (1+z) R^2} \left({\nu_a'\over \nu_i'}\right)^{1/3},
\label{nua1}
\end{equation}
where primes refer to quantities in the source comoving frame.  To
convert to the observer frame, we use
\begin{equation}
 \nu_a\equiv \nu_a' \Gamma/(1+z).
\end{equation}
Combining the above two equations we find
\begin{equation}
\nu_a = 3.8 f_i^{3/5} \nu_i^{-1/5} \Gamma^{3/5} \gamma_i^{-3/5} 
   R_{15}^{-6/5} (1+z)^{-9/5} d_{L28}^{6/5}.
\end{equation}
As before, all frequencies are in eV and fluxes in Jy.  Substituting
for $\Gamma$ \& $\gamma_i$ from equations (\ref{gam1}) and
(\ref{gami1}) results in
\begin{equation}
\nu_a = 5.1 f_i^{3/4} \nu_i^{-7/20} (\delta t)^{-3/10} R_{15}^{-6/5}
   Y^{-3/20} (1+z)^{-9/5} d_{L28}^{3/2}.
\label{nua2}
\end{equation}
As we discuss in \S3.2, we can estimate from the gamma-ray data the
ratio $\eta$ of the synchrotron frequency $\nu_i$ to the
self-absorption frequency $\nu_a$:
\begin{equation}
\eta \equiv \nu_i/\nu_a.
\label{eta1}
\end{equation}
Therefore, this is a third observational constraint on the source
properties.

To summarize, the source of optical emission is described by means of
five parameters.  The pulse duration $\delta t$, the optical flux
$\fop$ and the frequency ratio $\eta$ (eq. \ref{eta1}) give three
constraints.  The solution space is thus reduced to a two-dimensional
surface.  Additional constraints are obtained from the gamma-ray data,
as we discuss next.

\subsection{Gamma-ray emission via the inverse-Compton process}

We assume that the gamma-rays are produced by inverse Compton (IC)
scattering of synchrotron photons.  The peak frequency $\nu_{ic}$ of
the IC spectrum and the flux $f_{ic}$ at the peak are related to
$\nu_i$ and $f_i$ as follows,
\begin{equation}
\nu_{ic} \approx 3 \gamma_i^2 \nu_i, \quad\quad f_{ic} \approx 3\tau_e f_i
        = 3 Y \gamma_i^{-2} f_i,
\end{equation}
where a multiplicative factor of 3 in the expression for $f_{ic}$
takes into account the ratio of solid-angle integrated specific
intensity inside the source and the flux just outside the shell (in
the source comoving frame).  Substituting for $\gamma_i$ from equation
(\ref{gami1}),
\begin{eqnarray}
&& \nu_{ic6} = 1.9\times10^{-2} f_i^{-1/2}\nu_i^{3/2}Y^{1/2} R_{15}
    (1+z) d_{L28}^{-1}, \label{nuic1} \\
&& f_{ic-3} = 0.47 f_i^{3/2}\nu_i^{-1/2}Y^{1/2} R_{15}^{-1} (1+z)^{-1} 
    d_{L28},
\label{fic1}
\end{eqnarray}
where $\nu_{ic6} = \nu_{ic}/(10^6$ eV) and $f_{ic-3} =
f_{ic}/(10^{-3}$ Jy).

Using equation (\ref{nuic1}) for $\nu_{ic}$ we determine the synchrotron
peak flux,
\begin{equation}
f_i = 3.8\times10^{-4} \nu_{ic6}^{-2}\nu_i^{3} R_{15}^2 Y (1+z)^2 d_{L28}^{-2},
\label{fi2}
\end{equation}
and substituting this into equation (\ref{nua2}) we obtain $\nu_i$:
\begin{equation}
\nu_i = 1.2\times10^{2}\eta^{-10/9}\nu_{ic6}^{5/3}(\delta t)^{1/3} 
   R_{15}^{-1/3} Y^{-2/3} (1+z)^{1/3}.
\label{nui2}
\end{equation}
Eliminating $\nu_i$ from equation (\ref{fi2}) by using
eq. (\ref{nui2}), we find
\begin{equation}
f_i = 6.6\times10^2 \eta^{-10/3}\nu_{ic6}^{3}\delta t R_{15} Y^{-1} (1+z)^{3}
   d_{L28}^{-2}.
\label{fi3}
\end{equation}

Substituting for $\nu_i$ and $f_i$ from equations (\ref{nui2}) \&
(\ref{fi3}) into equation (\ref{fop0}) for the optical flux we obtain
\begin{equation}
\fop = \left\{ \begin{array}{ll}
\hskip -7pt 166\, \eta^{-{80\over 27}} \nu_{ic6}^{{22\over 9}} (\delta
t)^{{8\over 9}} R_{15}^{{10\over 9}} Y^{{-{7\over 9}}} (1+z)^{{26\over 9}}
d_{L28}^{{-2}}, & \nu_i>2, \\ & \\
\hskip -7pt 922\times8.7^{p}\, \eta^{-{5(p+6)\over9}} 
   \nu_{ic6}^{{5p+18\over6}} (\delta t)^{{p+6\over6}} R_{15}^{{6-p\over6}} Y^{-{p+3\over3}}
    (1+z)^{{p+18\over6}} d_{L28}^{-2}(\phi_\nu/\phi_f)^{{3+p\over3}}, 
    \quad\quad & \nu_i<2. 
\end{array} \right.
\label{fop1}
\end{equation}
We use the optical flux to eliminate one more unknown variable, $Y$,
\begin{equation} Y = \left\{ \begin{array}{ll}
\hskip -7pt 714\, \eta^{-{80\over21}} \fop^{-{9\over7}} \nu_{ic6}^{{22\over7}} 
   (\delta t)^{{8\over7}}     R_{15}^{{10\over7}} (1+z)^{{26\over7}} d_{L28}^{-{18\over7}},
     & \nu_i>2, \\
   & \\
\hskip -7pt 471\times1.4^{{3\over p+3}}\, \eta^{-{5(p+6)\over3(p+3)}}\fop^{-{3
   \over p+3}}
    \nu_{ic6}^{{5p+18\over2p+6}} (\delta t)^{{p+6\over2p+6}} R_{15}^{{6-p\over2p+6}} 
       (1+z)^{{p+18\over2p+6}} d_{L28}^{-{6\over p+3}}, \quad\quad & \nu_i<2.
\end{array} \right.
\label{y1}
\end{equation}
We are now in a position to express all quantities in terms of four
observables, viz., the pulse duration $\delta t$, the optical flux
$\fop$, the dimensionless self-absorption frequency $\eta$, and the
peak frequency of the gamma-ray spectrum $\nu_{ic}$, plus one unknown
parameter $R_{15}$.

The peak frequency of the synchrotron spectrum is obtained from equations 
(\ref{nui2}) \& (\ref{y1}):
\begin{equation} \nu_i = \left\{ \begin{array}{ll}
\hskip -7pt 1.5\, \eta^{{10\over7}} \fop^{{6\over7}} \nu_{ic6}^{-{3\over7}}
    (\delta t)^{-{3\over7}} R_{15}^{-{9\over7}} (1+z)^{-{15\over7}} d_{L28}^{{12\over7}},      & \nu_i>2, \\
   & \\
\hskip -7pt 2\times1.4^{-{2\over p+3}}\, \eta^{{10\over3(p+3)}}\fop^{{2\over p+3}}
     \nu_{ic6}^{-{1\over p+3}} (\delta t)^{-{1\over p+3}} R_{15}^{-{3\over p+3}} 
       (1+z)^{-{5\over p+3}} d_{L28}^{{4\over p+3}}, \quad\quad & \nu_i<2.
\end{array} \right.
\label{nui3}
\end{equation}
Let us define the transition radius $R_{tr}$ as that value of $R$
for which $\nu_i=2$~eV.  From equation (\ref{nui3}) we find
\begin{equation}
R_{tr,15} = 0.8 \eta^{{10\over9}} f_{op}^{{2\over3}}\nu_{ic6}^{-{1\over3}} 
   (\delta t)^{-{1\over3}} (1+z)^{-{5\over3}} d_{L28}^{{4\over3}}.
\label{rtr15}
\end{equation}
Note that, for $R>R_{tr}$, $\nu_i<2$~eV, and vice versa.

The flux at the peak of the synchrotron spectrum is obtained from
eqs. (\ref{fi3}) \& (\ref{y1}):
\begin{equation} f_i = \left\{ \begin{array}{ll}
\hskip -7pt 0.92\, \eta^{{10\over21}} \fop^{{9\over7}} \nu_{ic6}^{-{1\over7}}
    (\delta t)^{-{1\over7}} R_{15}^{-{3\over7}} (1+z)^{-{5\over7}} d_{L28}^{{4   \over7}},      & R<R_{tr}, \\
   & \\
\hskip -7pt 1.4^{{p\over p+3}}\, \eta^{-{5p\over3(p+3)}}\fop^{{3\over p+3}}
     \nu_{ic6}^{{p\over 2p+6}} (\delta t)^{{p\over 2p+6}} R_{15}^{{3p\over 2p+6}}
       (1+z)^{{5p\over 2p+6}} d_{L28}^{-{2p\over p+3}}, \quad\quad & R>R_{tr},
\end{array} \right.
\label{fi4}
\end{equation}
and the peak gamma-ray flux is obtained from equations (\ref{fic1}),
(\ref{y1}), (\ref{nui3}) \& (\ref{fi4}):
\begin{equation} f_{ic-3} = {\phi_\nu\over\phi_f}\left\{ \begin{array}{ll}   
 \hskip -7pt 12.3\,\, \eta^{-{40\over21}} \fop^{{6\over7}} \nu_{ic6}^{{11\over7}}
    (\delta t)^{{4\over7}} R_{15}^{-{2\over7}} (1+z)^{{6\over7}} d_{L28}^{-{2\over7}},
     & R<R_{tr}, \\ 
   & \\ \hskip -7pt 8.4\times1.4^{{2p+4\over p+3}}\,\, \eta^{-{10(p+2)\over3(p+3)}}
     \fop^{{2\over p+3}} \nu_{ic6}^{{2p+5\over p+3}} (\delta t)^{{p+2\over p+3}} 
      R_{15}^{{p\over p+3}} (1+z)^{{3p+4\over p+3}} d_{L28}^{-{2p+2\over p+3}}, 
   \quad\quad & R>R_{tr}.\end{array} 
\right.\label{fic2}
\end{equation}

Using equations (\ref{y1}), (\ref{nui3}) \& (\ref{fi4}) to substitute
for $Y$, $\nu_i$ \& $f_i$, equations (\ref{gami1}), (\ref{b1}) \&
(\ref{ne2}) give
\begin{equation} \gamma_i = \left\{ \begin{array}{ll}
\hskip -7pt 555\, \eta^{-{5\over7}} \fop^{-{3\over7}} \nu_{ic6}^{{5\over7}} 
   (\delta t)^{{3\over14}} R_{15}^{{9\over14}} (1+z)^{{15\over14}} d_{L28}^{-{6\over7}}, & R<R_{tr}, \\
   & \\
\hskip -7pt 527\, \eta^{-{5\over3(p+3)}}\fop^{-{1\over p+3}}
     \nu_{ic6}^{{p+4\over 2p+6}} (\delta t)^{{1\over 2p+6}} R_{15}^{{3\over 2p+6}}
      (1+z)^{{5\over 2p+6}} d_{L28}^{-{4\over 2(p+3)}}, \quad\quad & R>R_{tr},
\end{array} \right.
\label{gami2}
\end{equation}
\begin{equation} B = \phi_\nu^{-1}\left\{ \begin{array}{ll}
\hskip -7pt 4.5{\rm G}\,\, \eta^{{20\over7}} \fop^{{12\over7}} \nu_{ic6}^{-{13  \over7}}
    (\delta t)^{-{5\over14}} R_{15}^{-{43\over14}} (1+z)^{-{53\over14}} 
    d_{L28}^{{24\over7}}, & R<R_{tr}, \\
 & \\
\hskip -7pt 6.7{\rm G}\,\, \eta^{{20\over3(p+3)}}\fop^{{4\over p+3}}
     \nu_{ic6}^{-{p+5\over p+3}} (\delta t)^{{p-1\over 2p+6}} R_{15}^{-{p+15\over 2p+6}}
       (1+z)^{-{17-p\over 2p+6}} d_{L28}^{{8\over p+3}}, \quad\quad & R>R_{tr},
\end{array} \right.
\label{b2}
\end{equation}
\begin{equation} N_{e55} = \left\{ \begin{array}{ll}
\hskip -7pt 6.1\times10^{-3}\, \eta^{-{50\over21}} \fop^{-{3\over7}} 
      \nu_{ic6}^{{12\over7}} (\delta t)^{{5\over7}} R_{15}^{{15\over7}} 
      (1+z)^{{11\over7}} d_{L28}^{-{6\over7}}, & R<R_{tr}, \\
 & \\
\hskip -7pt 5.9\times10^{-3}\, \eta^{-{5(p+4)\over3(p+3)}}\fop^{-{1\over p+3}}
   \nu_{ic6}^{{3p+10\over 2p+6}} (\delta t)^{{p+4\over 2p+6}} R_{15}^{{3p+12\over 2p+6}}
      (1+z)^{{p+8\over 2p+6}} d_{L28}^{-{2\over p+3}}, \quad\quad & R>R_{tr}.
\end{array} \right.
\label{ne3}
\end{equation}

The energy in the magnetic field as measured in the GRB host galaxy
rest frame is $E_B = B^2 R^3/2$.  Using equation (\ref{b2}), this can
be shown to be
\begin{equation} E_B = \left\{ \begin{array}{ll}
\hskip -7pt 1.2\times10^{46}{\rm erg}\,\, \eta^{{40\over7}} \fop^{{24\over7}} 
    \nu_{ic6}^{-{26\over7}}
    (\delta t)^{-{5\over7}} R_{15}^{-{22\over7}} (1+z)^{-{53\over7}}
    d_{L28}^{{48\over7}}\phi_\nu^{-2}, & R<R_{tr}, \\
 & \\
\hskip -7pt 2\times10^{46}{\rm erg}\,\, \eta^{{40\over3(p+3)}}\fop^{{8\over p+3}}
     \nu_{ic6}^{-{2p+10\over p+3}} (\delta t)^{{p-1\over p+3}} R_{15}^{{2p-6\over p+3}}
       (1+z)^{-{17-p\over p+3}} d_{L28}^{{16\over p+3}}\phi_\nu^{-2},
   \quad\quad & R>R_{tr}.
\end{array} \right.
\label{eb}
\end{equation}
The energy in the charged particles ($e^\pm$) that produce the optical
and gamma-ray emission is $ E_{e} = N_e\gamma_i\Gamma m_e c^2$.
From equations (\ref{gam1}), (\ref{gami2}) \& (\ref{ne3}) this is
\begin{equation} E_{e} = 
   \left\{ \begin{array}{ll}
   \hskip -7pt 4\times10^{51}{\rm erg}\,\, \eta^{-{65\over21}} \fop^{-{6\over7}}
    \nu_{ic6}^{{17\over7}}
    (\delta t)^{{3\over7}} R_{15}^{{23\over7}} (1+z)^{{22\over7}}
    d_{L28}^{-{12\over7}}(\phi_\nu/\phi_f)^{{1\over2}}, & R<R_{tr}, \\
 & \\
\hskip -7pt 3\times10^{51}{\rm erg}\,\, \eta^{-{(5p+25)\over3(p+3)}}
    \fop^{-{2\over p+3}} \nu_{ic6}^{{2p+7\over p+3}} (\delta t)^{{1\over p+3}}
     R_{15}^{{2p+9\over p+3}} (1+z)^{{p+8\over p+3}} d_{L28}^{-{4\over p+3}} 
    (\phi_\nu/\phi_f)^{{1\over2}}, \quad\quad & R>R_{tr}.
\end{array} \right.
\label{epar}
\end{equation}
We should also consider the energy in the protons arising from the
bulk relativistic motion of the shell.  However, to estimate this
quantity we need to make some assumption regarding the composition of
the fluid in the shell, whether it is primarily an $e^+e^-$ plasma or
a $p^+e^-$ plasma.  In the former case the energy in protons is
negligible, while in the latter case the energy is $E_p=N_e\Gamma
m_pc^2 \sim {\rm few}\,\times E_e$ (taking $\gamma_i$ of order a few
hundred).

In the observer frame, the synchrotron cooling time $t_{syn}$ is
\begin{equation} 
t_{syn} = (7.7\times10^8 s) (1+z) \Gamma^{-1} B^{-2}\gamma_i^{-1},
\end{equation} 
which, using equations (\ref{gam1}), (\ref{gami2}) and (\ref{b2}),
can be written as
\begin{equation} t_{syn} ={\phi_\nu^{5/2}\over\phi_f^{1/2}}\left\{\begin{array}{ll}
   \hskip -7pt (620{\rm s})\,\, \eta^{-5} \fop^{-3} \nu_{ic6}^{3}
    (\delta t) R_{15}^{5} (1+z)^{7} d_{L28}^{-6}, & R<R_{tr}, \\
 & \\
\hskip -7pt (295{\rm s})\,\, \eta^{-{35\over3(p+3)}} \fop^{-{7\over p+3}}
   \nu_{ic6}^{{3p+16\over 2p+6}} (\delta t)^{{4-p\over 2p+6}} 
    R_{15}^{{p+24\over 2p+6}} (1+z)^{{32-p\over 2p+6}}
    d_{L28}^{-{14\over p+3}},     \quad\quad & R>R_{tr}.
\end{array} \right.
\label{tsyn}
\end{equation}
In addition to the loss of energy via synchrotron radiation, electrons
also lose energy through IC scattering of the local radiation field.
We calculate the photon energy density in the source rest frame from
the observed bolometric luminosity $L_{obs}$, and use this to
estimate the IC cooling time $t_{ic}$ in the observer frame:
\begin{equation}
 t_{ic} = {4\pi R^2 \Gamma (1+z) m_e c^2\over L_{obs}\gamma_i\sigma_T} = 
    (0.2{\rm s}) L_{obs,52}^{-1} (\delta t)^{-1/2} R_{15}^{5/2} 
     (1+z)^{3/2} \gamma_i^{-1}.
\end{equation}
Strictly speaking we do not know the true bolometric luminosity, so
the above estimate of $t_{ic}$ is an upper limit to the actual 
inverse-Compton cooling time. Using equation (\ref{gami2}) we can 
rewrite $t_{ic}$ in the following more useful form:

\begin{equation} t_{ic} = {\phi_\nu^{{1\over2}}\over\phi_f^{{1\over2}}}
   \left\{ \begin{array}{ll}
   \hskip -7pt {4\times10^{-4}{\rm s}\over L_{obs,52}}\,\, \eta^{{5\over7}} 
    \fop^{{3\over7}} \nu_{ic6}^{-{5\over7}}
    (\delta t)^{-{5\over7}} R_{15}^{{13\over7}} (1+z)^{{3\over7}}
    d_{L28}^{{6\over7}}, & R<R_{tr}, \\
 & \\
\hskip -7pt {4.5\times10^{-4}{\rm s}\over L_{obs,52}}\,\, \eta^{{5\over3(p+3)}}
    \fop^{{1\over p+3}} \nu_{ic6}^{-{p+4\over 2p+6}} (\delta t)^{-{p+4\over 2p+6}}
     R_{15}^{{5p+12\over 2p+6}} (1+z)^{{3p+4\over 2p+6}} d_{L28}^{{2\over p+3}},
    \quad\quad & R>R_{tr}.
\end{array} \right.
\label{tic}
\end{equation}

All the results obtained so far are general and could be applied to
any GRB that has the required data.  We now consider the implications
for the naked eye burst GRB 080319B.

\section{Application to GRB 080319B: Ruling out the internal shock model}


The relevant observational parameters for GRB 080319B are: $z=0.94$,
$d_{L28}=1.9$, $\fop=10$~Jy, $\nu_{ic6} = 0.665$, $p=5$, and $\delta
t\sim1$~s (from the gamma-ray variability).  Moreover, we estimate
that\footnote{The spectral indices are obtained by fitting the data
with the Band function, which gives the asymptotic value for the low
energy and high energy index. It should be noted that the IC spectrum
below $\sim 2\gamma_i^2\nu_a$ is $f_\nu\propto \nu$ and between this
frequency and the peak of $\nu f_\nu$ at $3\gamma_i^2\nu_i$ the
spectrum changes from $f_\nu\propto\nu$ to $\nu^{-1}$.  Therefore,
somewhere in between these two frequencies the index would be
$0.2$. The Konus data for GRB 080319B found the spectral peak to be at
665 keV and the observations extended down to a minimum photon energy
of 20~keV. Therefore, the IC spectral index of 0.2 at 20 keV requires
$\nu_i/\nu_a\sim 25$.} the time-averaged $\eta\equiv \nu_i/\nu_a$ for
this burst was about 25 (because $f_\nu\propto\nu^{0.18 \pm0.01}$
between 20keV and 650 keV), whereas the initial value of $\eta$ was
$\sim10$ as $f_\nu\propto\nu^{0.5\pm0.04}$ during the first 8\,s.
Scaling all quantities to these values, the transition radius $R_{tr}$
becomes
\begin{equation}
 R_{tr,16} = 12 \eta_{1.4}^{{10\over9}} f_{op,1}^{{2\over3}} \nu_{ic5.8}^{-{1\over3}}
    (\delta t)^{-{1\over3}}.
\label{rtr16}\end{equation}
If $R_{16} > R_{tr,16}$, then $\nu_i < 2$~eV and the optical band is
in the steep decaying part of the synchrotron spectrum, above the
synchrotron peak.  If $R_{16} < R_{tr,16}$, then the synchrotron peak
is above 2~eV, and the optical band is in the $F_\nu \propto
\nu^{1/3}$ part of the synchrotron spectrum.  Since the prompt optical
emission in GRB 080319B was exceptionally bright, it is likely that
the peak of the synchrotron spectrum was fairly close to the optical
band.  This suggests that $R_{16}$ must be within a factor of a few of
$R_{tr,16}$.  According to equation (\ref{rtr16}), $R_{tr} \sim
10^{17}$~cm, which is orders of magnitude larger than the radius at
which internal shocks are expected.  In fact, it is comparable to the
deceleration radius of the jet.

From the results described in \S3, we obtain the following numerical
results for the relevant parameters in GRB 080319B:
\begin{equation}
\Gamma = 572 (\delta t)^{-1/2} R_{16}^{1/2},
\label{gam08}
\end{equation}
\begin{equation} \gamma_i = \left\{ \begin{array}{ll}
\hskip -7pt 77\, \eta_{1.4}^{-{5\over7}} f_{op,1}^{-{3\over7}}\nu_{ic5.8}^{{5\over7}} 
   (\delta t)^{{3\over14}} R_{16}^{{9\over14}}, \quad\quad\quad & R<R_{tr}, \\
   & \\
\hskip -7pt 251\, \eta_{1.4}^{-{5\over24}}f_{op,1}^{-{1\over8}}
     \nu_{ic5.8}^{{9\over 16}} (\delta t)^{{1\over 16}} R_{16}^{{3\over 16}},
      \quad\quad\quad\quad\quad\quad & R>R_{tr},
\end{array} \right.
\label{gami08}
\end{equation}
\begin{equation} N_{e} = \left\{ \begin{array}{ll}
   \hskip -7pt (1.1\times10^{51}\,\, \eta_{1.4}^{-{50\over21}}
    f_{op,1}^{-{3\over7}} \nu_{ic5.8}^{{12\over7}}
    (\delta t)^{{5\over7}} R_{16}^{{15\over7}}, & R<R_{tr}, \\
 & \\
  \hskip -7pt (3.4\times10^{51}\,\, \eta_{1.4}^{-{15\over8}}
     f_{op,1}^{-{1\over8}} \nu_{ic5.8}^{{25\over16}} (\delta t)^{{9\over16}}
     R_{16}^{{27\over16}}, \quad\quad & R>R_{tr},
\end{array} \right.
\label{ne08}
\end{equation}
\begin{equation} Y = \left\{ \begin{array}{ll}
\hskip -7pt 2.5\times10^{-3}\, \eta_{1.4}^{-{80\over21}} f_{op,1}^{-{9\over7}} 
    \nu_{ic5.8}^{{22\over7}}(\delta t)^{{8\over7}} R_{16}^{{10\over7}},
     & R<R_{tr}, \\
   & \\
\hskip -7pt 7.5\times10^{-2}\, \eta_{1.4}^{-{55\over24}}f_{op,1}^{-{3\over 8}}
    \nu_{ic5.8}^{{43\over16}} (\delta t)^{{11\over16}} R_{16}^{{1\over16}},
       \quad\quad & R>R_{tr}.
\end{array} \right.
\label{y08}
\end{equation}
\begin{equation} E_{B} = \left\{ \begin{array}{ll}
   \hskip -7pt (2\times10^{55}{\rm erg})\,\, \eta_{1.4}^{{40\over7}} 
    f_{op,1}^{{24\over7}} \nu_{ic5.8}^{-{26\over7}}
    (\delta t)^{-{5\over7}} R_{16}^{-{22\over7}}, & R<R_{tr}, \\
 & \\
  \hskip -7pt (2.3\times10^{51}{\rm erg})\,\, \eta_{1.4}^{{5\over3}}
     f_{op,1}^{{}} \nu_{ic5.8}^{-{5\over2}} (\delta t)^{{1\over2}} 
     R_{16}^{{1\over2}}, \quad\quad\,\, & R>R_{tr},
\end{array} \right.
\label{eb08}
\end{equation}
\begin{equation} E_{e} = \left\{ \begin{array}{ll}
   \hskip -7pt (3.4\times10^{49}{\rm erg})\,\, \eta_{1.4}^{-{65\over21}} 
     f_{op,1}^{-{6\over7}} \nu_{ic5.8}^{{17\over7}}
    (\delta t)^{{3\over7}} R_{16}^{{23\over7}}, & R<R_{tr}, \\
 & \\
  \hskip -7pt (3.3\times10^{50}{\rm erg})\,\, \eta_{1.4}^{-{50\over24}}
          f_{op,1}^{-{1\over4}} \nu_{ic5.8}^{{17\over8}} (\delta t)^{{1\over8}} 
          R_{16}^{{19\over8}}, \quad\quad\,\, & R>R_{tr},
\end{array} \right.
\label{epar08}
\end{equation}
\begin{equation} t_{syn} = \left\{ \begin{array}{ll}
   \hskip -7pt (7.4\times10^{-4}{\rm s})\,\, \eta_{1.4}^{{-5}}
      f_{op,1}^{-3} \nu_{ic5.8}^{3} (\delta t) R_{16}^{5}, & R<R_{tr}, \\
 & \\
  \hskip -7pt (2{\rm s})\,\, \eta_{1.4}^{-{35\over24}} f_{op,1}^{-{7\over8}} 
     \nu_{ic5.8}^{{31\over16}}(\delta t)^{{1\over16}} R_{16}^{{29\over16}}, \quad\quad\quad\quad\quad\quad\quad\quad & R>R_{tr},
\end{array} \right.
\label{tsyn08}
\end{equation}
\begin{equation} t_{ic} = \left\{ \begin{array}{ll}
   \hskip -7pt {2.1{\rm s}\over L_{obs,52}}\,\, \eta_{1.4}^{{5\over7}} 
     f_{op,1}^{{3\over7}} \nu_{ic5.8}^{-{5\over7}} (\delta t)^{-{5\over7}} 
    R_{16}^{{13\over7}}, & R<R_{tr}, \\
 & \\
  \hskip -7pt {0.7{\rm s}\over L{obs,52}}\,\, \eta_{1.4}^{{5\over24}} 
         f_{op,1}^{{1\over8}} \nu_{ic5.8}^{-{9\over16}} (\delta t)^{-{9\over16}}
         R_{16}^{{37\over16}}, \quad\quad\quad\quad\quad\quad & R>R_{tr},
\end{array} \right.
\label{tic08}
\end{equation}
\begin{equation} f_{ic-3} = \left\{ \begin{array}{ll}
   \hskip -7pt 7.1\times10^{-2}\,\, \eta_{1.4}^{-{40\over21}} f_{op,1}^{{6\over7}}
      \nu_{ic5.8}^{{11\over7}} (\delta t)^{{4\over7}} R_{16}^{-{2\over7}},  & R<R_{tr}, \\
 & \\
  \hskip -7pt 7.8\times10^{-3}\,\, \eta_{1.4}^{-{35\over12}} 
       f_{op,1}^{{1\over4}} \nu_{ic5.8}^{{15\over8}} (\delta t)^{{7\over8}} 
       R_{16}^{{5\over8}}, \quad\quad\quad\quad\quad & R>R_{tr}.
\end{array} \right.
\label{fic08}
\end{equation}

\begin{figure}
\includegraphics[width=5.7in]{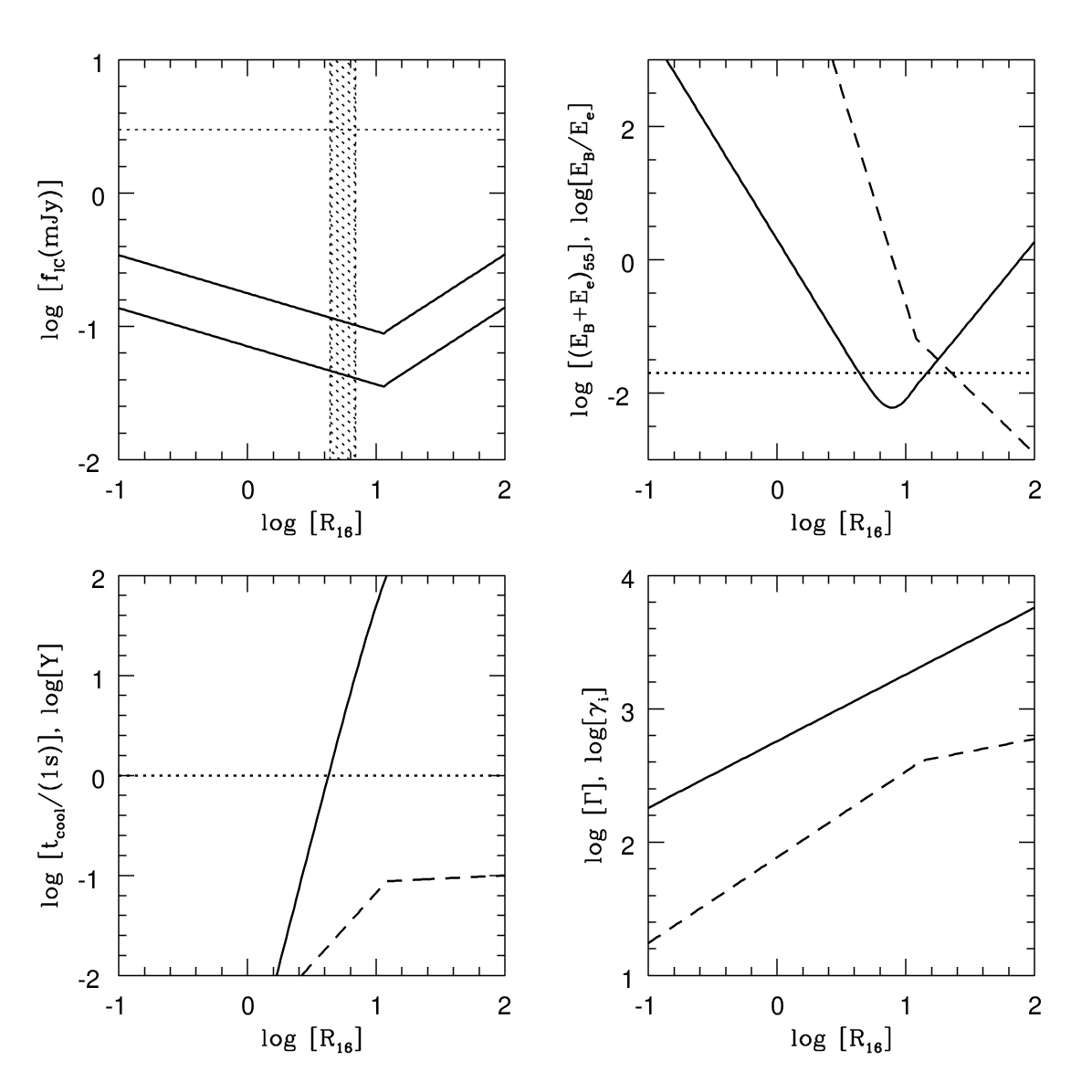}
\caption
{Top left: Shows the IC flux in GRB 080319B at 650~keV predicted by
the internal shock model. The lower line is from eq. (\ref{fic08}) and
the upper line corresponds to a factor of 2.5 larger flux to allow for
the expansion of the source during a photon crossing time, which was
not included in calculations presented in \S3 and \S4 (see the
Appendix for details and for a discussion of uncertainties in $f_{ic}$
and $\nu_a$). The shaded band is the region of the source radius
$R_{16}$ that is favored by various constraints (see text for
details).  For these values of $R_{16}$, the predicted IC flux falls
short of the observed gamma-ray flux (horizontal dotted line) by a
factor $\gta30$.  Top right: The solid line shows the total isotropic
energy in a single variability spike in the gamma-ray lightcurve.  The
dotted line shows the maximum energy allowed. The dashed line shows
$E_B/E_e$.  Bottom left: The solid line shows the cooling time in
units of the variability time (1\,s) -- which should be $\gta\ 1$
since the low energy spectral index was positive -- and the dashed
line shows the Compton $Y$.  Bottom right: The solid line shows the
bulk Lorentz factor $\Gamma$ and the dashed line shows the typical
Lorentz factor of electrons $\gamma_i$.}
\label{intshock}
\end{figure}

Figure 1 shows the dependences of a number of quantities as functions
of the only free parameter in the model: $R_{16}$.  The lower solid
line in the top left panel shows the predicted gamma-ray flux
$f_{ic-3}$ (based on eq. \ref{fic08}) and immediately indicates a
major problem.  If, as we suggested earlier, $R_{16}\sim R_{tr,16}$,
then the peak IC flux $f_{ic}$ that the model predicts falls short of
the observed flux by nearly a factor of 100

In Appendix A we discuss possible sources of error in our estimate of
the IC flux.  We show that the uncertainty in $f_{ic}$, even after
allowing for inhomogeneities in the source, is no larger than a factor
of order unity. The largest error is that we have overestimated
$\nu_a$ by a factor $\sim1.5$ by not including the expansion of the
source during the time it takes for a photon to cross the shell (see
Appendix A). The effect of this is that $f_{ic}$ is underestimated by
a factor $\sim2.5$ due to its dependence on $\nu_a$ via $\eta$.  Even
after correcting for this (upper solid line in upper left panel in
Fig.~1), the theoretically calculated gamma-ray flux is still smaller
than the observed value by a factor $\sim 30$.  This discrepancy is
much too large to be overcome by minor adjustments to the model.  We
thus conclude that the internal shock model with $R_{16}\sim
R_{tr,16}$ is ruled out for GRB 080319B.

One way to mitigate this problem is to select values of $R_{16}$ that
are either very much smaller or very much larger than $R_{tr,16}$.
However, as Fig.~1 shows, we need to modify $R_{16}$ by a huge factor,
which immediately leads to other problems.

The top right panel indicates one of the problems we face.  This panel
shows the total isotropic energy of the source $(E_B+E_e)$ in units of
$10^{55}~{\rm erg}$.  We see that shifting $R_{16}$ substantially away
from $R_{tr,16}$ causes the total energy to become unphysically large.
A reasonable upper limit to the total energy is
$10^{55}$\,erg\footnote{The isotropic energy release for GRB
080319B in the 20\,keV --- 7\,MeV band was 1.3x10$^{54}$\,erg. The
radiative efficiency for GRBs varies widely from burst to burst but is
generally larger than $\sim 10$\% (Panaitescu \& Kumar,
2002). Therefore, the total energy release in GRB 080319B is expected
to be $\lta\ 1.3$x10$^{55}$erg, and so we take $(E_B+E_e)\ \lta\
10^{55}$.}, which corresponds to an energy of
$\sim2\times10^{53}$\,erg in each spike in the gamma-ray lightcurve
The energy estimates in equations (\ref{eb08}) and (\ref{epar08})
refer to the latter and the limit is shown by the dotted line in the
top right panel.  We see that the energy constraint restricts $R_{16}$
to lie within the range $4.4-14.6$.  Within this range of $R_{16}$, we
have approximate equipartition between $E_B$ and $E_e$ (see the dashed
line), which is desirable, whereas choosing other values of $R_{16}$
would cause large deviations from equipartition.

The bottom left panel in Fig.~1 shows another set of problems.  Given
the huge luminosity of GRB 080319B, we expect the source to be
radiatively efficient, which means that $t_{\rm cool}$ must be
comparable to the variability time $t_{\rm var} \sim \delta t \sim
1$~s.  We see that the cooling time $t_{\rm cool}$, calculated
according to
\begin{equation}
{1\over t_{\rm cool}} = {1\over t_{\rm syn}}+{1\over t_{\rm ic}},
\end{equation}
is within a factor of 10 of the variability time only for models with
$R_{16}$ in the narrow range $2.7-6.9$.  (The variation of $t_{\rm
cool}$ with $R_{16}$ is extremely steep, so the condition $t_{\rm
cool}\sim t_{\rm var}$ is very restrictive.)  Combining this
constraint with the one we obtained earlier from the total energy, the
allowed range of $R_{16}$ is limited to $4.4-6.9$, shown as the
hatched vertical band in the top left panel in Fig.~1.  We also see
from the bottom left panel that the Compton $Y$ is less than 0.1 for
most models, and extremely small, $Y\ll 0.1$, for small values of
$R_{16}$.  Since $Y$ determines the fraction of the source luminosity
that comes out in gamma-rays, and since GRB 080319B (indeed, any GRB)
is a strong gamma-ray source, it seems unlikely that $Y$ could be this
small.

Finally, the bottom right panel in Fig.~1 shows the dependence of the
bulk Lorentz factor $\Gamma$ and the random electron Lorentz factor
$\gamma_i$ on $R_{16}$.  Models with $R_{16}\sim R_{tr,16}$ predict
reasonable values $\sim10^2-10^3$ for both Lorentz factors, but models
with very different values of $R_{16}$ predict either unusually low or
unusually high values.

In summary, all the indications suggest that the optical and gamma-ray
radiation in GRB 080319B were produced at a radius $R\sim {\rm
few}\,\times10^{16}\,{\rm cm}-10^{17}\,{\rm cm}$.  But at this radius,
the internal shock model predicts a negligibly small gamma-ray flux.
We are thus forced to conclude that the internal shock model, at least
in its standard form, is definitely ruled out for GRB 080319B.

\subsection{Other versions of the internal shock model}

We now consider whether we can get around the above difficulty by
modifying the internal shock model.  We begin by noting that, as long
as the gamma-ray emission is IC -- something that is required by the
low energy spectrum $f_\nu\propto \nu^{0.5}$ at early times (\S2) --
and the seed synchrotron photons are produced in the same source as
the $\gamma$-ray photons, equation (\ref{fic08}) is valid.  This
equation predicts an unacceptably low flux in the gamma-ray band.
Therefore, if we wish to save the internal shock model, we must give
up the assumption that all the radiation came from the same region of
the source.

Let us assume that the seed photons for IC scattering are produced by
the same source that gave us the optical flash. We will call this the
{\it optical region} of the source.  Let us assume that these seed
photons are IC-scattered in a different region, the {\it gamma-ray
region}.  We now show that the electron Lorentz factors $\gamma_i$ in
two regions are very similar.

Let us suppose that $\gamma_i$ in the optical region differs from that
in the gamma-ray region.  Then, the self-IC radiation from the optical
region will introduce a second IC component in the observed spectrum,
with a peak at a different photon energy.  Equation (\ref{fic08}) is
valid for any SSC process, so we can use it to estimate the flux in
the second peak. If the IC peak from the optical region is at a higher
photon energy than 650\,keV by a factor $> 2.5$, then equation
(\ref{fic08}) shows that the flux in this component will be larger
than the observed flux (note that $f_{ic}\propto\nu_{ic}^{11/7}$ as
per eq. \ref{fic08}, and the observed flux above 650 keV declined as
$\nu^{-2.87}$).  On the other hand, if the self-IC radiation peaks at
an energy much less than 650\,keV, the magnetic energy in the source
will increase very rapidly ($E_B\propto\nu_{ic}^{-26/7}$,
eq. \ref{eb08}). Since the energy is already close to the maximum
limiting value we can accept, this option is also ruled out.

Therefore, the values of $\gamma_i$ in the optical and gamma-ray
regions must be nearly the same.  This tight relation between the
Lorentz factors in the two regions suggests that the optical and
gamma-ray sources are very likely the same region. Even if they are
not, the similarity of their parameters means that the large
discrepancy in the gamma-ray flux discussed previously will survive
unchanged.  A related idea is that there are two populations of
electrons with different values of $\gamma_i$ within the same
source. One population is responsible for the seed photons and the
other for the IC scattering. This possibility can be ruled out by the
same argument.

This leads us to consider a model in which part of a shell is
magnetized -- this is where optical photons are produced -- and the
rest has a much weaker magnetic field (in order to avoid overproducing
synchrotron flux) but contains about 30 times more electrons in order
to produce the observed $\sim 3$mJy $\gamma$-ray flux via IC
scattering.  This situation can arise, for instance, when magnetic
field decays downstream of a shock front, as suggested in Kumar \&
Panaitescu (2008).  However, this proposal suffers from serious
problems that these authors have pointed out in their paper.  The
solution requires magnetic field to decay on a length scale that is
about 5\% of the shell thickness or about 10$^7$ plasma skin depth .
This scale corresponds to no particular physical scale in the system
and is quite arbitrary.  An even more severe problem is that the model
cannot account for the shorter time scale variability of gamma-rays
compared to the optical; in fact, the natural expectation is the
opposite in this model.

Note that Fig. 1 indicates an extremely narrow range of $R$ for the
radiating medium.  It is hard to believe that a large number of
independent shells ejected from the central source would all collide
at exactly this radius.  In addition, as we noted earlier, the radius
$R$ of the source is uncomfortably large for the internal shock model.
Both of these features would be explained naturally if we assumed that
the internal shocks are not between independent shells, but rather
between successive shells and the outermost shell, which is
decelerating after colliding with the external medium.  This is a
variation on the general idea of internal shocks (with a strong hint
of the forward shock model, see \S5.2.2), which at least provides an
explanation for the radius of the source.  However, this model can be
ruled out for two reasons. As with all other variants, this model
cannot explain the magnitude of the IC flux unless the magnetic field
occupies a small fraction of the shocked shell, about 5\% of the
ejecta width or 10$^7$ plasma skin depths. Furthermore, it predicts
that the pulse-width should increase with time, which is inconsistent
with the observed data for GRB 080319B which show, if anything, that
the last few pulses in the gamma-ray lightcurve were somewhat narrower
than the initial few pulses.

Having considered these and other ideas, we believe that it is
impossible to explain the observations of GRB 080319B with any
reasonable version of the internal shock model.  Fortunately, there is
an alternative model which invokes relativistic turbulence in the
radiating fluid.  We now apply this model to GRB 080319B.

\section{Relativistic turbulence model for GRB 080319B}

The basic kinematic features of the relativistic turbulence model are
described in Narayan \& Kumar (2008).  In brief, this model explains
the observed variability in GRB lightcurves by postulating an
inhomogeneous relativistic velocity field in the GRB-producing medium
(which we refer to as the ``shell'' because of its shell-like
morphology in the host galaxy frame).  The beaming effect of the
turbulent eddies causes large amplitude fluctuations in the observed
flux.  Despite being inhomogeneous, the model is radiatively efficient
in the sense that the whole medium radiates and the observer receives
a fair share of the radiated luminosity.  This important feature,
which is a direct consequence of beaming, allows the model to overcome
the arguments of Sari \& Piran (1997) against inhomogeneous GRB
models.  The reader is referred to Narayan \& Kumar (2008) for
details.

Since the relativistic turbulence model has a natural explanation for
the observed variability, equation (\ref{tau}) relating the
variability time scale $\delta t$ to $R$ and $\Gamma$ is no longer
needed.  Instead, the quantity $R/2\Gamma^2c$ determines the {\it
total} burst duration $t_\gamma$.  We thus have
\begin{equation}
t_\gamma \sim {R(1+z)\over 2c\Gamma^2} = (1.7\times10^4\,{\rm s})R_{15}
\Gamma^{-2}(1+z).\label{tgamma}
\end{equation}
Since $t_\gamma\sim50$~s for GRB 080319B, whereas $\delta t \sim
t_{var}\sim 1$~s, this modification has a rather profound effect on
the results.

In the relativistic turbulent model, we consider turbulent eddies with
a typical bulk Lorentz factor $\gamma_t$ in the frame of the shell,
and a typical size $\sim R/(\gamma_t\Gamma)$ in the comoving frame of
an eddy.  The eddies are volume-filling, so there are $\sim\gamma_t^3$
eddies in a causally connected region of volume $\sim R^3/\Gamma^3$.
We assume that the velocity field of eddies changes direction by $\sim
2\pi$ on the light crossing time scale $\sim R/(c\gamma_t\Gamma)$. In
this case the probability that an eddy, some time during its life,
will move towards the observer with a velocity vector within an angle
$(\gamma_t\Gamma)^{-1}$ of the line-of-sight is $\sim\gamma_t^{-1}$
(Narayan \& Kumar, 2008). Therefore, over the course of the burst, a
given observer will receive emission from $\gamma_t^2$ eddies, with
each eddy producing a pulse of radiation lasting a time (see Narayan
\& Kumar 2008 for details).
\begin{equation}
t_{var}\sim t_\gamma/\gamma_t^2.
\end{equation} 
Since GRB 080319 has $t_{var}\sim t_\gamma/100$, we infer that
$\gamma_t\sim10$ for this burst.  Note that, at any given time, the
observer receives radiation from only one eddy on average.

We assume that the fluid in the shell consists of eddies and an
inter-eddy medium.  The latter is produced when eddies collide and
shock.  Let us take the thermal Lorentz factor of electrons within an
eddy to be $\gamma_{it}$.  The thermal Lorentz factor of electrons in
the inter-eddy medium follows from energy conservation when eddies
collide, and is $\sim \gamma_{it}\gamma_t\equiv\gamma_i$. Similarly,
if we take the magnetic field in the inter-eddy frame to be $B$, then
the comoving magnetic field in an eddy is $B/\gamma_t^{1/2}$, assuming
that the magnetic energy is roughly conserved when eddies dissipate.
Using these scalings we see that the peak of the synchrotron spectrum
(as measured in the shell frame) for inter-eddy and eddy emissions are
proportional to $B\gamma_i^2$ and $B \gamma_i^2 \gamma_t^{-3/2}$,
respectively.

Let us take the average number of electrons in an eddy to be $N_{ed}$,
and the total number of electrons in the inter-eddy medium in a volume
$(R/\Gamma)^3$ (the volume of a causally connected region) to be
$N_i$.  For simplicity, let us assume that the total number of
electrons in all the $\gamma_t^3$ eddies is of order $N_i$, i.e., half
of the fluid in the shell is in eddies and the other half is in the
inter-eddy medium.  Thus we have $N_{ed}\sim N_i/\gamma_t^3$.

At any given time, only one eddy will produce beamed radiation towards
the observer.  The peak synchrotron flux from this eddy is
proportional to $\sim B N_{ed} \gamma_t^{3/2}\Gamma^3 \sim B
N_i\Gamma^3/\gamma_t^{3/2}$.  Here we have made use of the fact that,
at a fixed observer time, the observer receives radiation from only a
fraction of the electrons in the eddy, $\sim N_e/\gamma_t$, due to the
time dependence of eddy velocity direction.  The peak synchrotron flux
from the inter-eddy medium is $\sim B N_i\Gamma^3$, which is larger
than the peak flux from the eddy by a factor $\sim\gamma_t^{3/2}$.
The synchrotron flux in a fixed observer band above the peak frequency
is larger for the inter-eddy medium by an additional factor of
$\gamma_t^{3(p-1)/4}$.  We thus conclude that the synchrotron emission
observed in the optical band is completely dominated by the inter-eddy
medium.  We therefore ignore eddies when we estimate the optical
synchrotron flux.

The situation is different for the IC emission.  Let us write the
synchrotron flux as seen by a typical electron in the inter-eddy
medium as $f_{syn}$ (this is easily estimated from the calculation
above). The observed IC luminosity due to electrons in the inter-eddy
medium $f_{ic}^{ie}$ is then 
\begin{equation}
f_{ic}^{ie}\propto \sigma_T f_{syn} N_i \Gamma^3,
\label{ficie}
\end{equation}
while the IC emission from an eddy pointing towards the observer
$f_{ic}^{eddy}$ is
\begin{equation}
f_{ic}^{eddy}\propto \sigma_T (\gamma_t f_{syn}) N_{ed} 
(\Gamma\gamma_t)^3/\gamma_t \sim \sigma_T f_{syn} N_i \Gamma^3
\label{ficeddy}
\end{equation}
We see that the two contributions are
equal. Therefore, both components in the shell fluid contribute
equally to the gamma-ray IC flux.  Of course, the inter-eddy
contribution will vary smoothly over the duration of the burst,
whereas the contribution from the eddies will be highly variable.

Using these results, we may easily estimate the values of various
parameters in GRB 080319B corresponding to the relativistic turbulence
model.  Note that, since the synchrotron radiation, or optical flux,
comes from the inter-eddy plasma, it satisfies the same equations as
derived in \S4.  Moreover, the IC flux has no dependence on the
Lorentz factor of turbulent eddies (eq. \ref{ficeddy}).  Therefore,
equations (\ref{rtr16})--(\ref{fic08}) may be directly used for the
relativistic turbulence model provided we replace $\delta t$ with the
burst duration $50$\,s, and take $N_e$, $E_B$, $E_e$ and $f_{ic}$ to
be two times larger than the values given by these equations (the
factor of two is to count both the eddies and the inter-eddy medium).

Setting $\delta t=50$\,s (eq. \ref{tgamma}), $f_{op}=10$\,Jy,
$\nu_{ic}=650$\,keV and $\eta_{1.4}=1$ (i.e., $\eta=
\nu_i/\nu_a\approx25$) into equation (\ref{rtr16}) we obtain the
transition radius
\begin{equation}
R_{tr} = 2\times10^{16}\,{\rm cm}.
\label{Rtrturb}
\end{equation}
The other parameters follow from equations
(\ref{gam08})--(\ref{fic08})
\begin{equation}
\Gamma = 81 R_{16}^{1/2},
\label{gam08b}
\end{equation}
\begin{equation} \gamma_i = \left\{ \begin{array}{ll}
\hskip -7pt 254\, R_{16}^{{9\over14}}, 
    \quad\quad\quad & R<R_{tr}, \\
   & \\
\hskip -7pt 349\, R_{16}^{{3\over 16}},
      \quad\quad\quad\quad\quad\quad & R>R_{tr},
\end{array} \right.
\label{gami08b}
\end{equation}
\begin{equation} N_{e} = \left\{ \begin{array}{ll}   
  \hskip -7pt 9.4\times10^{52}\,\, R_{16}^{{15\over7}} & R<R_{tr}, \\
   & \\  \hskip -7pt 1.3\times10^{53}\,\, R_{16}^{{27\over16}}, 
   \quad\quad\quad\quad\quad\quad\quad\quad & R>R_{tr},\end{array} \right.\label{ne08b}
\end{equation}
\begin{equation} Y = \left\{ \begin{array}{ll}
\hskip -7pt 2.1  R_{16}^{{10\over7}}, & R<R_{tr}, \\
   & \\
\hskip -7pt 5.5\, R_{16}^{{1\over16}},
       \quad\quad\quad\quad\quad\quad & R>R_{tr},
\end{array} \right.
\label{y08b}
\end{equation}
\begin{equation} E_{B} = \left\{ \begin{array}{ll}
   \hskip -7pt (2.6\times10^{53}{\rm erg})\,\, R_{16}^{-{22\over7}}, & R<R_{tr}, \\
 & \\
  \hskip -7pt (1.9\times10^{52}{\rm erg})\,\, 
     R_{16}^{{1\over2}}, \quad\quad\quad\quad\quad\quad\, & R>R_{tr},
\end{array} \right.
\label{eb08b}
\end{equation}
\begin{equation} E_{e} = \left\{ \begin{array}{ll}
   \hskip -7pt (1.3\times10^{51}{\rm erg})\,\, R_{16}^{{23\over7}}, & R<R_{tr}, \\
 & \\
  \hskip -7pt (2.6\times10^{51}{\rm erg})\,\, 
          R_{16}^{{19\over8}}, \quad\quad\quad\quad\quad\quad\, & R>R_{tr},
\end{array} \right.
\label{epar08b}
\end{equation}
\begin{equation} t_{syn} = \left\{ \begin{array}{ll}
   \hskip -7pt (0.3{\rm s})\,\, R_{16}^{5}, & R<R_{tr}, \\
 & \\
  \hskip -7pt (2.8{\rm s})\,\, 
      R_{16}^{{29\over16}}, \quad\quad\quad\quad\quad\quad\quad\quad\quad\quad\quad & R>R_{tr},
\end{array} \right.
\label{tsyn08b}
\end{equation}
\begin{equation} t_{ic} = \left\{ \begin{array}{ll}
   \hskip -7pt {0.1{\rm s}\over L_{obs,52}}\,\, R_{16}^{{13\over7}}, & R<R_{tr}, \\
 & \\
  \hskip -7pt {0.07{\rm s}\over L{obs,52}}\,\, 
         R_{16}^{{37\over16}}, \quad\quad\quad\quad\quad\quad\quad\quad\quad\quad & R>R_{tr},
\end{array} \right.
\label{tic08b}
\end{equation}
\begin{equation} f_{ic-3} = \left\{ \begin{array}{ll}
   \hskip -7pt 0.80 R_{16}^{-{2\over7}},  & R<R_{tr}, \\
 & \\
  \hskip -7pt 0.44 R_{16}^{{5\over8}}, \,\quad\quad\quad\quad\quad\quad\quad\quad\quad\quad\quad\quad & R>R_{tr}.
\end{array} \right.
\label{fic08b}
\end{equation}
As discussed above, we include only the inter-eddy medium for
calculating the synchrotron component of the emission, but we include
both the eddies and the inter-eddy medium when calculating the IC
component. The synchrotron cooling time given by equation
(\ref{tsyn08b}) applies only to electrons in the inter-eddy medium;
the timescale is larger by a factor of $\gamma_t$ for electrons in
eddies.  We note that the IC flux should be increased by a factor of
2.5 to allow for the expansion of the source shell (Appendix A).
Also, the flux will be larger by a factor $N_{ed}\gamma_t^3/N_i$ than
given in equation (\ref{fic08b}) if there are more electrons in eddies
than in the inter-eddy medium.

\subsection{Relativistic turbulence: a consistent model for GRB 080319B}

\begin{figure}
\includegraphics[width=6.5in]{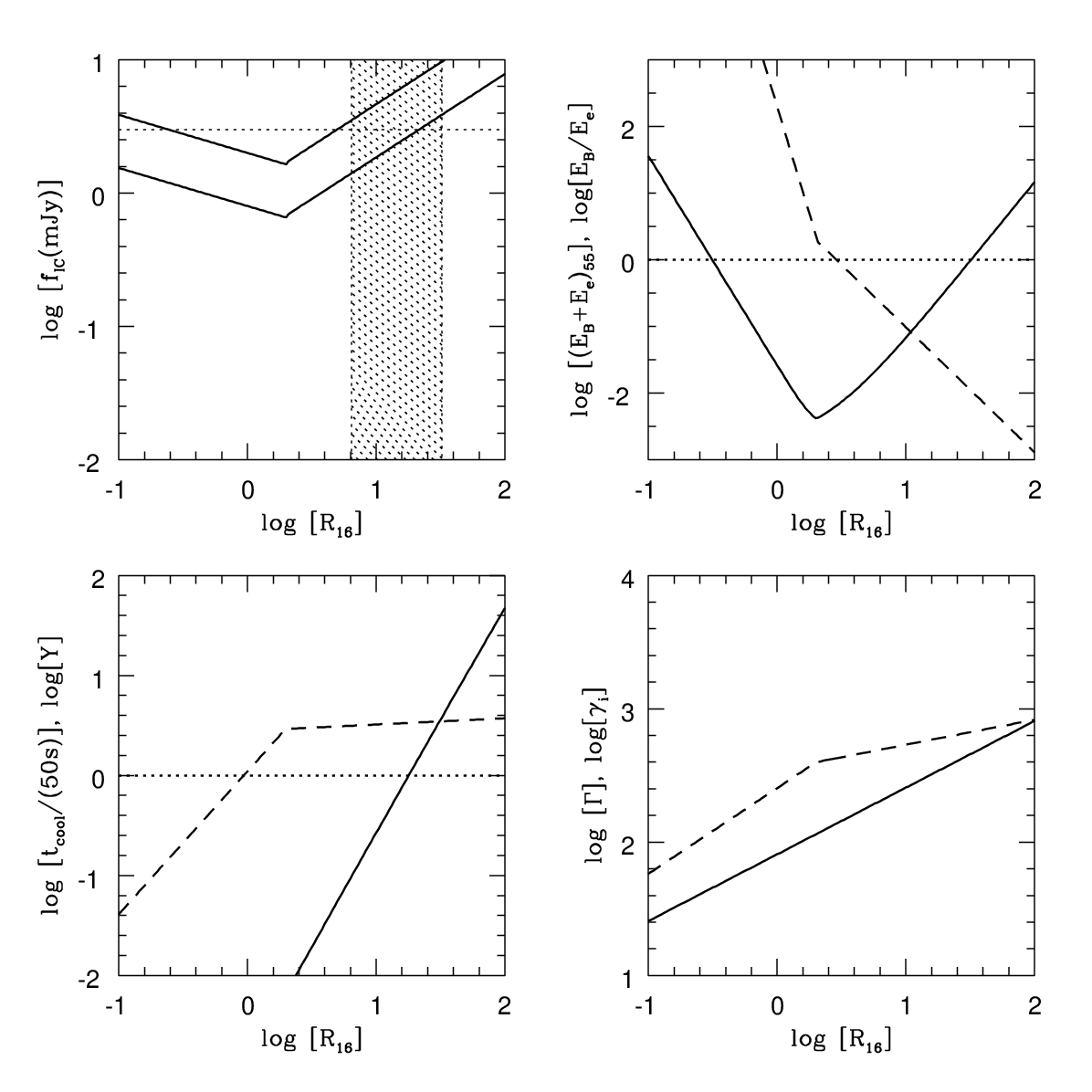}
\caption{Similar to Fig.~1, but for the relativistic turbulence model.
Note that the predicted IC flux (solid lines in the topic left panel)
is perfectly consistent with observations (horizontal dotted line).
The shaded region indicates the range of source radii $R_{16}$ that is
consistent with constraints on the total energy (solid line in the top
right panel) and the cooling time (solid line in the bottom left
panel) constraints (see text for details). We note that the ratio of IC 
and synchrotron luminosities is larger by $\sim 10$ than the value
of $Y$ shown as a dashed line in the lower left panel (see eq. \ref{Ypar}
for the definition of $Y$) due to two different effects each contributing 
a factor $\sim3$; (1) the commonly used Compton-$Y$ is larger than $Y$ 
in eq. \ref{Ypar} by a factor 3 due to $p$ dependent factors, and (2) the 
synchrotron photon energy density inside the shell is larger than the
naive estimate of $L_{syn}/(4\pi R^2 c)$ by a factor $\sim3$.  }
\label{turbmodel}
\end{figure}

Figure 2 is similar to Fig. 1, but shows what happens when we use the
relativistic turbulent model.  We find a very good match both with the
observations and with various consistency conditions when we choose
$R\approx 10^{17}$~cm. For this choice of $R$, we have (i) gamma-ray
flux $f_{ic-3}$ predicted to be close to the observed flux, (ii)
modest requirement for the total isotropic energy $\sim10^{54}$~erg,
(iii) $E_B\sim E_e/10$, i.e., approximate equipartition between
magnetic and thermal energy, (iv) $t_{\rm cool} \sim 50$~s, i.e., the
cooling time is comparable to the burst duration and thus consistent
with efficient radiation, (v) $Y\sim$~few\footnote{The ratio of energies
in the IC radiation and the synchrotron emission is $\sim 10Y$.
This is in part because the parameter $Y$, defined in
eq. \ref{Ypar}, differs from the commonly used Compton-$Y$ parameter by 
a factor $\sim3$ (due to $p$-dependent factors not included
in eq. \ref{Ypar}), and the mean synchrotron photon energy density in the
shell is larger than $L_{syn}/(4\pi R^2 c)$ by a factor 3.},
 i.e., consistent with
gamma-rays dominating the emission, (vi) $\Gamma\sim250$ as inferred
for GRBs in general from a variety of observations (Lithwick \& Sari,
2001), and (vii) electron Lorentz factor in the inter-eddy medium
$\gamma_i\sim500$ and in eddies $\gamma_i/\gamma_t\sim50$, which are
quite reasonable.  

The shaded band in Fig. 2 shows the range of $R_{16}$ that is
consistent with our two primary constraints.  First, we require the
total isotropic energy over the duration of the burst to be no larger
than $10^{55}$\,erg.  This constrains $R_{16}$ to lie in the range
$0.31-32$.  Second, we require the cooling time $t_{\rm cool}$ to lie
within a factor of 10 of the burst duration 50\,s.  This gives the
constraint $6.5 < R_{16} < 50$.  Requiring both conditions to be
satisfied simultaneously restricts $R_{16}$ to lie in the range
$6.5-32$, as shown in Fig. 2.  Within this range, the predicted
gamma-ray flux agrees remarkably well with observations.  

Note that the deceleration radius for the blast wave is
\begin{equation}
R_d = \left( {3E\delta t \over 4\pi \bar{n} m_p c (1+z) } \right)^{1/4}
   = 1.1\times10^{17}{\rm cm}\,\, E_{54}^{1/4} (\delta t/50s)^{1/4}
      \bar{n}^{-1/4},
\end{equation}
where $\bar{n}$ is the mean particle density of the circumstellar
medium within the radius $R_d$.  It is interesting that $R_d$ lies in
the middle of the allowed range for the source distance $R$. It provides
independent confirmation that the prompt radiation in GRB~080319B was
not produced in internal shocks -- there is no reason why internal
shocks should occur at the deceleration radius.

In addition to the various successes described above, the relativistic
turbulence model explains all the major qualitative features observed
in the $\gamma$-ray and optical lightcurves of GRB 080319B during the
initial $\sim 10^2$\,s, i.e., before the onset of forward shock
emission.

Since the gamma-ray emission (via IC) arises partly from eddy
electrons and partly from the inter-eddy medium, we expect the
gamma-ray lightcurve to consist of a smooth slowly-varying component
plus a large number of sharp spikes.  This is the case for most GRBs,
including GRB 080319B (e.g., Fig. 1 in Racusin et al. 2008).  The
relative fluxes in the two components provide information on the
relative numbers of electrons in the two media.  We assumed in our
model (for convenience) that the numbers are roughly equal and this is
reasonably consistent with the observations.  As already mentioned, by
combining the $\gamma$-ray variability time of $\sim 0.5$s with the
burst duration of 50s, we infer that $\gamma_t\sim10$ for GRB 080319B.

Since the synchrotron emission is generated by inter-eddy electrons,
the optical light curve is expected to be much less variable than the
IC-dominated gamma-ray emission.  This is indeed the case for GRB
080319B.  At the same time, the overall duration of the optical and
gamma-ray lightcurves are expected to be similar, as observed.

The optical lightcurve of GRB 080319B showed an initial rapid rise by
more than an order of magnitude in flux (Fig. 3 in Racusin et
al. 2008), whereas the $\gamma$-ray lightcurve showed a much less
rapid rise.  This finds a natural explanation. According to the
turbulent model of GRB 080319B, the synchrotron frequency $\nu_i$ was
below the optical band.  Therefore, the optical spectrum is predicted
to be very soft: $F_\nu \propto \nu^{-2.8}$.  If we assume that
$\nu_i$ initially started off at a somewhat lower frequency and later
settled down at a larger value, say by a factor of $\sim 3$, then the
optical flux would increase by nearly a factor of 20.  This
explanation might indicate that the gamma-ray peak energy $\nu_{ic}$
should also increase with time, whereas in fact $\nu_{ic}$ decreased
by a small amount (from 750~keV to 550~keV).  To explain this, we
would need to invoke that the electron Lorentz factor $\gamma_i$
decreased by a factor of about 2 during this time.  Note that the
reason for the much less rapid increase of the $\gamma$-ray flux is
that the synchrotron peak flux, which is proportional to the number of
electrons and the magnetic field strength, is a slowly varying
function of time.

Another property of the relativistic turbulence model is that we
should continue to see emission in the $\gamma$-ray band for a time
duration somewhat longer than the prompt optical lightcurve
duration. The reason is that there is a very high probability that a
few eddies lying a little bit outside of $\Gamma^{-1}$ will point
toward the observer, thereby slightly lengthening the burst duration in the
gamma-ray band (see Fig. 1 in Narayan \& Kumar, 2008). This effect is
clearly seen in the lightcurves of GRB 080319B; the optical LC started
falling off at 43s whereas the steep decline of the gamma-ray LC 
began at 51s. 

A prediction of the model is that the synchrotron and IC spectra
should be the same. In particular, since our solution for GRB 080319B
requires $R\sim10^{17}\,{\rm cm}>R_{tr}$, or $\nu_i<2$\,eV, the spectral
index in the optical band during the burst should have been the same
as the high energy index in the gamma-ray band, i.e. $\beta=2.87$.  It
is unfortunate that there were no measurements of optical spectrum
during the burst.  The first measurement was at $t\sim10^2$\,s when it
was found that $\beta=0.55$ or $f_\nu \propto\nu^{-0.55}$ (Wozniak et
al. 2008). This measurement would seem to call into question the
prediction of the SSC model.

It is, however, interesting to note that the optical lightcurve showed
a sharp break at about 90\,s. Prior to this time the flux scaled as
$t^{-5.5}$ and after this time the flux decreased as $t^{-2.8}$ (see
Fig. 2 of Kumar \& Panaitescu or Racusin et al. 2008).  This suggests
that 90\,s marked a transition from one source of radiation to
another, and that $\beta=0.55$ at $\sim10^2$s corresponds to the
second source which gave rise to the $f_\nu\propto t^{-2.8}$ part of
the lightcurve and possibly unrelated to the prompt radiation.  

The optical lightcurve decline of $t^{-5.5}$ between 43s and 90s is
roughly consistent with the expectation of the relativistic turbulence
model after the source is turned off at $t\sim43$s; the observed
radiation in this case is the large-angle emission (LAE) from photons
arriving from angles larger than $\Gamma^{-1}$, leading to a flux
decline of $t^{-2-\beta}$ or $\sim t^{-5}$ when $\beta\sim3$ (Kumar \&
Panaitescu, 2000). So the steeply declining optical lightcurve at the
end of the GRB provides an indirect confirmation of a steep spectrum
in the optical band.  We note that the temporal behavior of flux from
an adiabatically expanding shell of angular size $\Gamma^{-1}$ is
similar to the large angle emission. The decay index of the lightcurve
for adiabatic expansion is somewhat steeper than LAE for a given
$\beta$ (Barniol-Duran \& Kumar, 2008), and therefore this is a
preferred mechanism for the observed optical flux during 43--90s.

The gamma-ray lightcurve in the 15-150 keV at the end of the burst was
seen to fall off even faster than the optical flux at the end of the
burst. This is a puzzling behavior, and unlikely to be due to LAE. The
reason is that the spectral index in this band was close to zero
during the burst, and therefore the LAE flux decline should be $\sim
t^{-2}$ --- unless the peak frequency fell off from 650 keV to less
than 100 keV at the end of the burst which seems unlikely. The only
natural explanation for the steep decline of the $\gamma$-ray flux is
that the angular size of the source was $\sim \Gamma^{-1}$, and
gamma-rays for $t>51$\,s were from the adiabatically cooling source;
the IC flux from an adiabatically cooling shell declines much faster
than the synchrotron lightcurve (Barniol-Duran \& Kumar, 2008).

What about the fall-off of the optical flux as $t^{-2.8}$ for $t\
\gta\ 10^2$\,s?  It cannot be LAE for the reasons described above.  We
offer an explanation for this part of the optical lightcurve that
requires the GRB jet to be a Poynting outflow (see \S5.2.4). A
Poynting jet traveling outward from the center of the star cannot
avoid sweeping up and accumulating some amount of baryonic material at
its head. In subsequent jet expansion this baryonic gas is cooled, and
at the deceleration radius it is heated once again by the reverse
shock.  The optical emission from this reverse-shock heated gas might
be responsible for the lightcurve for $10^2\ \lta\ t\ \lta\
10^3$\,s. We know from equation (\ref{nrs2}) below that
$N_e/N_{RS}\sim 10$ (assuming that the kinetic energy in the baryonic
head of the Poynting jet is of order the explosion energy). Therefore,
the optical flux from the reverse shock is a factor $\sim 10$ smaller
than the prompt gamma-ray source.  The more slowly declining reverse
shock flux took over from the very rapidly declining flux from the
early GRB tail at $t\sim10^2$\,s, and continued to dominate the
lightcurve until the even more slowly declining, but weaker, forward
shock optical emission took over at $t\sim10^3$\,s. We note that the
effect of a narrow jet, with opening angle $\sim \Gamma^{-1}$, is very
weak on the emergent lightcurve decay for a long period of time when
the jet is propagating in a medium with density falling off as
$r^{-2}$ (Kumar \& Panaitescu, 2000); the density in the circumstellar
medium of GRB 080319B is in fact inferred to be $r^{-2}$ by the late
time afterglow data, cf. Racusin et al. (2008), Kumar \& Panaitescu
(2008).

\subsection{Where exactly is the turbulent region located?}

As we have seen, the relativistic turbulence model gives robust
estimates for various source parameters such as the radius, bulk
Lorentz factor and energy of the shell, the number and typical Lorentz
factors of the radiating electrons, etc.  Using these results we now
attempt to infer where the radiating region is located within the
context of a dynamical model of GRBs.

\subsubsection{Not in internal shocks}

The internal shock model, including all reasonable variations, is
firmly ruled out, as we have discussed in \S4.  Inclusion of
relativistic turbulence within the context of this model will not
salvage the situation unless we take $\delta t$ to be the burst
duration (as shown in \S5). However, in that case we are dealing with
a situation in which the emission region is close to the deceleration
radius, which is no longer an internal shock.

The primary motivation for the internal shock model is to explain the
rapid variability observed in the gamma-ray lightcurves of GRBs (Sari
\& Piran 1997).  The relativistic turbulence model described in
Narayan \& Kumar (2008) has a completely different explanation for the
variability.  In particular, this model no longer needs to assume
equation (\ref{tau}), which is the key relation in the internal shock
model.  Therefore, we see no reason to retain the internal shock
picture.

\subsubsection{Not in the forward shock}

In the standard model of GRBs, the collision of the relativistic
ejecta with the external medium causes a pair of shocks to be
generated: a {\it forward shock} (FS) which is driven into the
external medium and a {\it reverse shock} (RS) which is driven into
the ejecta.

We can rule out the FS by considering the number of electrons we need
for producing the observed radiation.  From equation (\ref{ne08b}) we
see that the radiating region must have about $6\times 10^{54}$
electrons.  However, the number $N_{FS}$ of electrons/protons
processed in the FS must satisfy, by a simple energy argument,
\begin{equation}
N_{FS}\Gamma^2 m_p c^2 = E/2, \quad{\rm or}\quad N_{FS} = {2E(\delta t)\over
   m_p c (1+z) R} = 2\times10^{52} E_{54} (\delta t)_1 (1+z)^{-1} R_{16}^{-1}.
\label{nfs}
\end{equation}
For the particular case of GRB 080319B this gives $N_{FS}\sim
10^{52}$, which is smaller than the number of electrons needed by a
factor $\sim10^2$.  This is a large discrepancy, so we can discard the
FS as the location of the relativistic turbulence.

\subsubsection{Relativistic turbulence in the reverse shock?}

Could the relativistic turbulence be located in the RS?  Let the GRB
ejecta be composed of protons and electrons,
and let us take the Lorentz factor of the RS front with respect to the
unshocked ejecta to be $\Gamma_{RS}$.  By applying pressure
equilibrium across the contact discontinuity between the FS and RS
fluid, we find the number of electrons $N_{RS}$ that have been
processed through the RS to be
\begin{equation}
N_{RS} = N_{FS} {\Gamma\over \Gamma_{RS}}.
\label{nrs}
\end{equation}
Using equations (\ref{gam08b}) \& (\ref{nfs}), and the parameters for 
GRB 080319B, we find 
\begin{equation}
N_{RS} = 1.3\times10^{54} E_{55} R_{17}^{1/2} \Gamma_{RS,1}^{-1}.
\label{nrs1}
\end{equation}
Thus, the ratio of the number of electrons needed for
optical/gamma-ray radiation (eq. \ref{ne08b}) and $N_{RS}$ is given by
\begin{equation}
{N_e\over N_{RS}} = 5 E_{55}^{-1} \Gamma_{RS,1}^{-1} R_{17}^{19/16}.
\label{nrs2}
\end{equation}

If gamma-rays were to arise in the reverse shock then we expect
$E_{55}\sim 0.4$. The reason is that half the energy of the blast wave
is in the reverse shock at the deceleration radius, and this energy is
efficiently radiated when the cooling frequency is close to $\nu_i$,
as seems to be the case for GRB 080319B. Moreover, we presumably
require $\Gamma_{RS}\ \gta\ 10$ in order for the shocked gas to have a
turbulent $\gamma_t\sim10$.  The requirement that $t_{cool}\sim50$\,s
means that $R_{17}\sim1$.  Therefore, we find from the above equation
that $N_e/N_{RS}\sim 10$.  This ratio might be closer to unity
provided that protons carry a much larger fraction of the blast wave
energy, so that $E_{55}$ is $\sim1-2$ rather than $0.4$.

The interesting result that it is possible to have $N_e\sim N_{RS}$
suggests that the turbulence is perhaps produced in the RS-heated GRB
ejecta. The ratio of energies in magnetic fields and particle kinetic 
energy in this case is $\sim 0.1$ (fig. 2), which is similar to the value
derived for the Crab pulsar at the wind termination shock (Kennel \& Coroniti,
1984). Presumably, the turbulence is a natural consequence of a
relativistic shock.  For instance, the contact discontinuity surface
separating the FS and RS region is known to suffer from the
Rayleigh-Taylor instability. Could this explain the turbulence?  In
the shell comoving frame, the growth rate of the Rayleigh-Taylor
instability at the interface of a relativistic RS and FS can be shown
to be
\begin{equation}
\omega^2 = f k g, 
\end{equation}
where $g\sim c^2\Gamma/R$ is the effective gravitational acceleration
in the shell comoving frame, $k=2\pi \ell \Gamma/R$ is the wavenumber
of the perturbation, $f=3/(8\Gamma_{RS}^2 - 5)$, and $\Gamma_{RS}$ is
the Lorentz factor of the RS front with respect to the unshocked
ejecta.  For $\Gamma_{RS}\gg1$ the above equation reduces to
\begin{equation}
\omega \sim {c\Gamma\ell^{1/2}\over \Gamma_{RS} R} \sim {\ell^{1/2}\over\Gamma_{RS}
  \delta t' },
\end{equation}
where $ \delta t'$ is the GRB duration in the shell comoving frame.
Thus, the number of e-folds by which the Rayleigh-Taylor mode can grow
is $\sim \omega (\delta t')\sim \ell^{1/2} \Gamma_{RS}^{-1}$. The eddy
scale $\ell$ of interest to the IC problem is $R/(\Gamma\Gamma_{RS})$
or $\ell\sim \Gamma_{RS}$. Perturbations on this scale will undergo
$\Gamma_{RS}^{-1/2}$ e-folds of growth, i.e., the amplitude increases
by less than a factor 2.  Therefore, the Rayleigh-Taylor instability
is not sufficiently potent to generate the highly relativistic
turbulence we need.

Recently Goodman \& MacFadyen (2007) and Milosavljevic, Nakar \& Zhang (2007)
have discovered interesting instabilities, resulting from a clumpy 
circumstellar medium and an initially anisotropic blastwave respectively,
which lead to vorticity generation downstream of the shock front. These
instabilities have been further studied by Sironi \& Goodman (2007),
and Milosavljevic et al. (2007) to investigate the generation 
of magnetic fields in relativistic shocks. Couch, Milosavljevic \& Nakar
(2008) have found another instability that generates vorticity down
stream of a shockfront even when the circumstellar medium is homogeneous
and the blastwave isotropic. We have estimated the growth 
rate of these instabilities and find that these too fail to give rise 
to relativistic turbulence.

Of course, we cannot rule out the possibility that there might be
other as yet unknown instabilities that might give rise to
relativistic turbulence.  Therefore, we are unable to discard the
possibility that the prompt GRB emission originates in the RS.

\subsubsection{Relativistic turbulence in the Poynting-dominated jet}

A Poynting-dominated jet would have a weak reverse shock (Kennel \&
Coroniti, 1984; Zhang \& Kobayashi, 2005) and would not be consistent
with the proposal considered in the previous subsection.  On the other
hand, such a jet probably undergoes various plasma instabilities at
the deceleration radius.  These instabilities would stir up the fluid
into a state consistent with our model of relativistic turbulence.
The instabilities would presumably heat up the electrons until
quasi-equipartition is achieved, consistent with the results shown in
Fig. 2.

According to equations (\ref{eb08b}) \& (\ref{epar08b}), in our model
$E_B/E_e\sim 0.1$ at $R\sim10^{17}$\,cm.  However, this does not rule
out the Poynting outflow model. The reason is that in all of our
formulae $B$ really stands for the projection of the magnetic field
vector perpendicular to the electron momentum vector i.e.,
$B\sin\alpha$ where $\alpha$ is the pitch angle between the electron
momentum and the magnetic field direction. For a random distribution
of particle pitch angle the difference between $B$ and $B\sin\alpha$
is order unity. However, when electrons have a non-zero average
momentum along the local magnetic field (as might be the case for
particles accelerated in reconnection regions), the difference can be
large.  For instance, when the average $\alpha$ is 0.3 the energy in
magnetic fields is larger than that in equation (\ref{eb08b}) by a
factor $\sim10$, making the model consistent with equipartition.

Lyutikov and Blandford (2003) have suggested that the dissipation
of magnetic energy in a Poynting flux dominated jet should occur at a
distance of $\sim 3\times 10^{16}$cm due to current driven
instabilities (see Lyutikov 2006 for a concise summary of the model,
and for a comparison with the baryonic outflow model).  Acceleration
of electrons (and positrons), and plasma bulk flow along the magnetic
field lines at roughly the local Alfven speed are expected in the
process of magnetic field decay/reconnection. These expectations of
the Poynting outflow model are roughly consistent with our findings
for GRB 080319B: emission generated at $R\approx10^{17}$cm and
turbulent velocity field with Lorentz factor $\gamma_t\sim
10$. However, the reason for a very soft particle spectrum, $p\sim5$,
is unclear (at least to us); numerical simulations of particle
acceleration in reconnection regions generally find a hard particle
spectrum (e.g., Larrabee et al. 2003).

Moreover, it is also not clear how $\gamma_t$ and $\gamma_i$ should be
related to the bulk $\Gamma$ of the pre-instability jet. Nor is it
clear why the typical Lorentz factor of electrons should be a modest
value $\gamma_i\approx500$ with the kind of powerful accelerator one
might expect in magnetic reconnections (other than the fact that it is
energetically impossible to accelerate a large number of electrons, of
order $10^{55}$, to an average Loretnz factor much larger than $\sim
500$). Further investigation is required to address these questions.

\section{Summary}

We have shown in this paper that the gamma-ray and optical data for
GRB 080319B rule out the popular internal shock model for generation
of the prompt radiation.  According to this model, the duration
($\lta\ 1$\,s) of spikes in the gamma-ray lightcurve sets an upper
bound on the quantity $R/(2c\Gamma^2)$, where $R$ is the radius of the
source relative to the center of the explosion and $\Gamma$ is the
bulk Lorentz factor.  When we apply this condition, we find that it is
impossible to fit the observed optical and gamma-ray flux
simultaneously.  Specifically, any model that fits the optical flux
under predicts the gamma-ray flux by nearly two orders of magnitude
(Fig.~1).  This is an unacceptably large discrepancy which cannot be
eliminated with any reasonable modification of the internal shock
model.

An equally powerful qualitative argument against the internal shock
model is the fact that we find the radius $R$ of the source to be
constrained quite tightly by the observations.  The energy required in
magnetic field increases very rapidly as we decrease $R$, $E_B\propto
R^{-22/7}$, whereas the energy in particles increases rapidly,
$E_e\propto R^{17/7}$.  Also, the cooling time of electrons becomes
too short to be compatible with observations if
$R<3\times10^{16}$\,cm.\footnote{Collisions at a smaller radius would
produce a weak optical flash with flux decreasing roughly as
R$^{11/12}$.  The electrons would undergo very rapid cooling and
produce a low energy spectrum in the gamma-ray band of
$f_\nu\propto\nu^{-1/2}$. Prompt optical observations of GRB 080319B
show variations in the optical flux by less than a factor two for much
of the 50\,s duration of the burst except at the beginning and the
end.  Moreover, the low energy spectral index for the gamma-ray
emission was greater than 0 throughout the burst.}  All of these
factors together constrain the location where the prompt emission in
GRB~080319B was produced to lie within a narrow range of radius:
$4\times10^{16}\ \lta\ R\ \lta\ 8\times10^{16}$\,cm (see Fig.~1).
This is problematic for the internal shock model.  According to this
model, there is a large number of internal shocks among independent
ejecta, with a separate shock producing each of the $\sim50$ spikes in
the gamma-ray lightcurve of GRB 080319B.  Why would all the ejecta
collide within such a narrow range of radius?  Moreover, why should
the radius be so close to the deceleration radius $R_d\sim
10^{17}$\,cm, where the ejecta meet the external medium and begin to
slow down?  This coincidence is suspicious.

All of these problems are eliminated if we give up the internal shock
model and consider instead a model in which the variability in the
gamma-ray lightcurve is produced by relativistic turbulence in the
source with random eddy Lorentz factors $\gamma_t\sim10$.  In this
model, the quantity $R/(2c\Gamma^2)$ is no longer constrained to be
less than 1\,s, but only needs to be comparable to the burst duration
$\sim50$\,s (Narayan \& Kumar 2008).  With this modification, we find
that we obtain a remarkably consistent model of GRB 080318B (see
Fig.~2) in which the prompt optical emission was produced by
synchrotron emission and the gamma-rays were the result of inverse
Compton scattering.  The predicted gamma-ray flux is perfectly
compatible with observations.  Also, estimates of various quantities
such as the total energy, cooling time, Lorentz factor, etc. are all
very reasonable and consistent (\S5.1).  The radius of the source is
calculated to be in the range $6\times10^{16}<R<3\times10^{17}$\,cm;
if we select a nominal value $R\sim10^{17}$\,cm, we obtain an
excellent fit to all the observations.

In the context of a physical model, the picture that emerges from this
model is that the energy of the relativistic jet in GRB~080319B was
converted to optical \& $\gamma$-ray radiation either via a
relativistic reverse shock when ejecta (composed of $p^+$s and $e^-$s)
ran into the circumstellar medium or that much of the jet energy was
in magnetic field that was dissipated close to the deceleration
radius.  Theoretically, it is difficult to understand how a reverse
shock might produce relativistic turbulence with $\gamma_t\sim10$
(\S5.2.3).  Also, it is easier to understand the optical data for the
time period $10^2\ \lta\ t\ \lta\ 10^3$\,s if we assume a Poynting jet
(\S5.2.4).  For these reasons we have a mild preference for the
Poynting-dominated jet model.

A potential problem for the Poynting jet model is that the ratio of
magnetic to particle kinetic energy is about 0.1 for our best solution
(Fig.~2). However, this ratio is similar to that inferred for the
pulsar wind termination shock for the Crab pulsar (Kennel \& Coroniti,
1984). Moreover, this ratio of 0.1 does not rule out the Poynting model
for another reason which is
that, in all of our formulae, $B$ is the projection of the magnetic
field perpendicular to the electron momentum vector.  Thus, if
electron momenta are preferentially parallel to the magnetic field,
then the true $E_B$ would be larger than our estimate (easily by a
factor 10 compared to the value given in eq. \ref{eb08b}), and we can
have $E_B/E_{e}\sim1$.  Note that electrons are accelerated parallel
to the magnetic field in reconnection regions and so this possibility
is not as arbitrary as it might appear.

In the relativistic turbulence model, fluctuations in the observed
gamma-ray lightcurve are produced as a result of random relativistic
variations in the velocity field of the source, with turbulent Lorentz
factor $\gamma_t\sim 10$.  The model predicts that there should be
$\sim\gamma_t^2\sim100$ spikes in the gamma-ray light curve (Narayan
\& Kumar 2008), which is consistent with the $\sim50$ spikes seen in
GRB~080319B.  The optical synchrotron flux is dominated by the
inter-eddy medium rather than eddies.  Therefore, we expect much less
variability in the optical flux, as was indeed observed.

The model can explain the sharp rise in the optical flux of
GRB~080319B at the beginning of the burst.  For this, we must
postulate that the synchrotron peak frequency increased from
$\sim0.5$\,eV to $\sim1.5$\,eV during the first $\sim15$s.  Since the
synchrotron peak frequency $\nu_i$ was below the optical band (2\,eV),
the optical flux was smaller than the peak synchrotron flux by a
factor $(4/\nu_i)^{2.8}$ (the spectral index above the peak is known
from gamma-ray observations).  A modest increase in $\nu_i$ by a
factor of 3 early in the burst would thus produce a factor of 20
increase in the observed optical flux.  The reason that the gamma-ray
flux increased by a much smaller factor during the same time is that
the peak IC flux is proportional to $\tau_e N_e B$, which would
change little during this time period.

The end of the gamma-ray prompt emission phase occurred $\sim 8$\,s
after the prompt optical in GRB~080319B.  This is probably a result of
inverse Compton emission from turbulent eddies lying a bit outside of
the primary $1/\Gamma$ cone, but which happened to point toward us
because of a fortuitous alignment of their turbulent velocity (the
probability for this happening is of order unity).  Since much of the
synchrotron emission comes from non-turbulent fluid in between eddies,
the optical flux would not have a similar effect.

\section*{Acknowledgments}
PK is grateful to Rodolfo Barniol Duran for checking all equations in this 
paper, and Milos Milosavljevic for a number of useful discussions regarding
relativistic turbulence.

\appendix

\section{Possible errors in the calculation of IC flux}

We discuss in this appendix possible sources of error in our calculation
of the IC flux, i.e. errors associated with eq. (\ref{fic08}); according
to this equation the theoretically calculated gamma-ray flux is smaller 
than the observed value by a factor $\sim 20$.

A possible source of error might arise from our assumption of a
homogeneous source, and we need to estimate its effect on the IC flux.
The synchrotron peak flux is a linear function of magnetic
field strength and the total number of electrons in the source, and 
therefore clumping of electrons and $B$, to lowest order, have little 
effect on the emergent flux. The IC flux is, however, affected by 
clumping of electrons, and we estimate the magnitude of this effect.

Let us consider an extreme form of inhomogeneity where all electrons
are concentrated in $M_c$ clumps of each size $r_c$. The number density
of electrons in the clumps is $n_c$, and the density averaged over the
source volume, $\sim R^3$, is $n_0$; there are no electrons in between
clumps. Let us assume that the synchrotron power from each electron is 
$p_\nu$. In this case the synchrotron luminosity of the source is 
\begin{equation}
L_{syn}(\nu) \approx p_\nu (n_c r_c^3) M_c \approx p_\nu n_0 R^3,
\end{equation}
and is independent of electron clumping. The IC luminosity depends
on synchrotron flux in the vicinity of electrons (in clumps). The  
synchrotron flux is
\begin{equation}
f_{syn}(\nu) \approx p_\nu (n_c r_c + n_0 R) \approx {L_{syn}(\nu)\over
  R^2} \left[ 1 + {n_c r_c\over n_0 R} \right].
\end{equation}
The IC luminosity is obtained from the above flux:
\begin{equation}
L_{ic}(\nu_{ic}) \approx \sigma_T f_{syn}(\nu) n_c r_c^3 M_c
 \approx \sigma_T L_{syn}(\nu) n_0 R \left[ 1 + {n_c r_c\over n_0 R} \right].
\end{equation}
Or
\begin{equation}
L_{ic}(\nu_{ic}) \approx \sigma_T L_{syn}(\nu) n_0 R \left[1 + 
   f_c^{-2/3} M_c^{-1/3} \right],
\end{equation}
where $f_c = n_c r_c^3 M_c/(n_0 R^3)$ is the fraction of the shell volume
occupied by clumps. We see from the above equation that IC flux can be
enhanced by clumping of electrons. For instance, consider an 
example where $f_c=0.1$ and $M_c=1$ (all electrons are in a single small
clump). The IC flux in this case is a factor 4.5 larger than when electrons
are uniformly distributed. This flux enhancement is about an order of 
magnitude smaller than what is needed to explain the observed flux for
GRB 080319B (eq. \ref{fic08}). 
An even more extreme case of clumping could bridge the gap, however the
efficiency for converting jet energy to radiation is very small 
when $f_c\ll1$ as pointed out by Sari and Piran (1997). Furthermore,
another serious problem is that a high degree of
clumping leads to an increase of $\nu_a$ (as discussed below), and that
makes the SSC spectrum below the peak inconsistent with the observed
data for GRB 080319B -- unless we place the source at a distance 
from the center of explosion that is larger than the deceleration radius.

The dependence of $f_{ic}$ on $\eta\equiv\nu_i/\nu_a$ is fairly strong
and so we need to discuss the uncertainty in $\eta$. 
We have taken $\eta=25$ ($\log\eta=1.4$), which is guided by the Konus-Wind
low energy spectrum of $f_\nu\propto\nu^{0.2}$ in the energy band
20--650 keV. This low energy spectral index suggests that
$\nu_a$ (the self-absorption frequency) should be $\lta20$ keV,
and thus $\eta \gta 32$; therefore, $\eta=25$ is a conservative choice for 
GRB 080319B. However, is it possible that $\nu_a$ has been overestimated
in our calculation by our assumption of a homogeneous source? If $\nu_a$ 
were to be smaller by a factor $\sim 6$ than given by equation (\ref{nua2}) 
then that would lead to a larger IC
flux by  factor 30 (see eq. \ref{fic08}), and thereby reconcile the 
observed and the theoretically expected gamma-ray flux. We show that 
inhomogeneities in the source cannot decrease $\nu_a$ as long as the 
optical flux we observed during the burst is produced in the source.

We calculate synchrotron self-absorption frequency ($\nu_a$) when 
$B$, $\gamma_i$ \& $n_0$ are allowed to vary, arbitrarily, across 
the source; the electron distribution is taken to be $dn/d\gamma = n_0 
(\gamma/\gamma_i)^{-p}$ for $\gamma\ge \gamma_i$. Spatial variations 
in $B$, $\gamma_i$ \& $n_0$ are subject to constraints
that the optical flux and the IC peak frequency should be
equal to the observed values.

Our starting point is equation (6.52) of Rybicki \& 
Lightman (1979) for the synchrotron absorption coefficient, $\alpha_\nu$; 
$\int dr'\, \alpha_\nu$ is the optical depth 
for absorbing synchrotron photons of frequency $\nu$.
We can show that for a power-law electron distribution
and for $\nu<\nu_i$ (the case of interest for 080319B)
\begin{equation}
\alpha_{\nu'} \approx {3^{1/2}(p+2)(p-1)q^3 n_0 B\sin\delta\over 16\pi^2(p+2/3)
    m_e{\nu'}^2\gamma_i} \left( {\nu'\over \nu'_i}\right)^{1/3},
\end{equation}
where
\begin{equation}
 \nu'_i \equiv {3 q B\sin\delta \,\gamma_i^2\over  4\pi m_e c},
\end{equation}
$\delta$ is the angle between magnetic field and electron velocity vector,
and prime denotes frequency in the source comoving frame.
The synchrotron self-absorption frequency is determined from the equation
\begin{equation}
\int dr'\, \alpha_{\nu'_a} \approx {3^{1/2}(p+2)(p-1)q^3 \over 16\pi^2(p+2/3)
    m_e{\nu'_a}^2} \int dr'\, {n_0 B\sin\delta\over \gamma_i} 
   \left( {\nu'_a\over \nu'_i}\right)^{1/3}=1.
\end{equation}
Or
\begin{equation}
{\nu'_a}^{5/3} \approx {3^{1/2}(p+2)(p-1)q^3 \over 16\pi^2 m_e (p+2/3) }
    \int dr'\, {n_0 B\sin\delta\over {\nu'_i}^{1/3}\gamma_i}.
\label{nua5}
\end{equation}
The $\nu'_a$ given by the above equation is self-absorption frequency
along one line of sight. Since an observer receives photons from an
area $\sim\pi R^2/\Gamma^2$, we should average $\nu_a$ over this
area. This average frequency is given by
\begin{equation}
\langle{\nu'_a}^{5/3}\rangle \approx {3^{1/2}(p+2)(p-1)q^3 \over 16\pi^2 m_e
    (p+2/3) } {\Gamma^2\over \pi R^2} \int d^3{\bf x}' \, {n_0 B\sin\delta\over 
    {\nu'_i}^{1/3} \gamma_i }.
\end{equation}

Since the optical flux ($f_{op}$) is proportional to $\int d^3{\bf x}'\, 
n_0 B\sin\delta/{\nu'_i}^{1/3}$, when $\nu'_i$ lies above the optical
band, it is convenient to define a new variable $\chi \equiv n_0 B\sin\delta/
 {\nu'_i}^{1/3}$, and rewrite the equation for $\nu'_a$ using this new variable
\begin{equation}
\langle{\nu'_a}^{5/3}\rangle \propto {1\over R^2} \int d^3{\bf x}' \,
    {\chi\over \gamma_i}.
\label{nua7}
\end{equation}
The minimum of $\nu'_a$ can be obtained by requiring that $\delta \nu'_a =0$
for an infinitesimal variation of $\chi({\bf x'})$, i.e.
\begin{equation} 
 \int d^3{\bf x}' \, {\delta\chi({\bf x'})\over \gamma_i({\bf x'})} = 0,
\label{pert1}
\end{equation}
subject to the condition that
\begin{equation}
  \int d^3{\bf x}' \, \delta\chi({\bf x'}) = 0.
\end{equation}
The variational integral implicitly assumes that we are solving for 
$\gamma_i({\bf x'})$ which can be an arbitrary function as long as
the IC spectral peak of the radiation emergent from the source matches
the observed value, i.e. $1\ll\gamma_i\ll\infty$.
Let us consider a special form of $\delta\chi({\bf x'})$ that is
nonzero in two spherical regions of infinitesimal radius centered
at ${\bf x'_1}$ and ${\bf x'}_2$. It is required that $\delta\chi({\bf x'}_1)
= - \delta\chi({\bf x'}_2)$ in order to satisfy the optical flux constraint.
Substituting this into equation (\ref{pert1}) leads to 
$\gamma_i({\bf x'}_1) = \gamma_i({\bf x'}_2)$. Since, ${\bf x'}_1$ and 
${\bf x'}_2$ were arbitrary points in the source, we conclude that
$\gamma_i$ does not vary across the source when $\nu_a'$ is minimized.
Therefore, we can take $\gamma_i$ outside of the integral in equation
(\ref{nua7}), and find see that the minimum value of $\langle\nu_a\rangle$ is
fixed by the observed optical flux. In other words, the assumption 
of a homogeneous source used in our calculations in \S3 \& \S4 gives the 
smallest possible value for $\eta$ or the largest IC flux. For 
a high degree of clumping of electrons in the source, we considered
above, $\nu_a$ is larger when our line of sight passes through a clump
and especially when only one clump lies on our sight line;
there is, however, little change
to $\langle{\nu_a}\rangle$ as we have shown above. In this case an observer 
will receive radiation from one clump at a time, and will find $\nu_a$ 
larger than the case of a homogeneous shell. A larger $\nu_a$ for a 
clumpy source goes in the opposite direction to what we need to 
increase the IC flux, and this largely reverses the gain to the IC flux 
found above (eq. A4).

We assumed in our derivation that $\nu_i'$ is above the optical
band. A similar proof for the minimum of $\nu_a'$ can be carried
out when $\nu_i'$ is below the optical frequency. Moreover, we ignored
variations of $\Gamma$ across the source. This approximation is 
justified since large variation in $\Gamma$ are smoothed out in
less than one dynamical time.

There is one effect that we have not included in our calculation which
lowers the value of the self-absorption frequency $\nu_a$ somewhat.
As synchrotron photons propagate outward they move though a medium
where the electron density is decreasing with time (due to the outward
expansion of the shell). As a result, a calculation based on a
stationary source overestimates the optical depth to Thomson
scattering and $\nu_a$ by factors of $\sim 2$ \& $\sim2^{3/5}$,
respectively.  This can be seen by considering a homogeneous shell
with electron density $n_0$ located at a distance $R_0$ from the
center of explosion, with a radial thickness $R_0/\Gamma^2$.  (The
comoving frame shell thickness of $R_0/\Gamma$ is obtained by
causality considerations, since $R_0/c\Gamma$ is the time elapsed in
the shell comoving frame.)  As seen by a lab frame observer, a photon
moving outward in the radial direction takes a time $2R_0/c$ to cross
the shell, and during this time the shell has moved to a larger radius
and the density has decreased. For a hot shell, the radial width too
increases with time, and the photon transit time is a bit larger
still. A straightforward calculation of the optical depth to Thomson
scattering when the density changes as $1/R^2$ shows that the shell
optical depth is about half of what it is for a stationary medium of
the same thickness and density. Substituting this optical depth into
equation (\ref{nua5}) for $\nu_a$ we see that the synchrotron-self
absorption frequency is smaller for an expanding shell by a factor of
$\sim 2^{3/5}$.

The net effect is that the parameter $\eta$ which is defined in
equation (\ref{eta1}) should be reduced by a factor of $2^{3/5}$.  In
the case of GRB 080319B, we estimated $\eta\sim25$ from the
observations.  We should use a smaller value $\eta\sim 16$ in our
formulae to obtain more accurate numerical estimates of quantities.
This causes the predicted gamma-ray flux to be increased by a factor
of 2.5.

\end{document}